\documentclass[final,leqno,onefignum,onetabnum]{siamltexmm}

\usepackage{amsmath}
\usepackage{mathrsfs}
\usepackage{amssymb}
\usepackage{stmaryrd}
\usepackage{dsfont}
\usepackage{enumerate}

\newtheorem{example}{Example}

\title{A theoretical framework for calibration in computer models: parametrization, estimation and convergence properties \thanks{Tuo's research is partially sponsored by the Office of Advanced Scientific Computing Research; U.S. Department of Energy, project No. ERKJ259  ``A mathematical environment for quantifying uncertainty: Integrated and optimized at the extreme scale.'' The work was performed at the Oak Ridge National Laboratory, which is managed by UT-Battelle, LLC under Contract No. De-AC05-00OR22725. Tuo's work is also supported by the National Center for Mathematics and Interdisciplinary Sciences, CAS and NSFC 11271355. Wu's research is supported by NSF DMS-1308424 and DOE DE-SC0010548.}}

\author{Rui Tuo\thanks{Academy of Mathematics and Systems Science, Chinese Academy of Sciences, Beijing, China 100190; and Computer Science and Mathematics Division, Oak Ridge National Laboratory, Oak Ridge, TN 37831
(\email{tuorui@amss.ac.cn}).}
\and C. F. Jeff Wu\thanks{School of Industrial and Systems Engineering, Georgia Institute of Technology, Atlanta, GA 30332
(\email{jeffwu@isye.gatech.edu}).}}

\begin{document}
\maketitle
\newcommand{\slugmaster}{%
\slugger{juq}{xxxx}{xx}{x}{x--x}}

\begin{abstract}
Calibration parameters in deterministic computer experiments are those attributes that cannot be measured or available in physical experiments. Kennedy and O'Hagan \cite{kennedy2001bayesian} suggested an approach to estimate them by using data from physical experiments and computer simulations. A theoretical framework is given which allows us to study the issues of parameter identifiability and estimation. We define the $L_2$-consistency for calibration as a justification for calibration methods. It is shown that a simplified version of the original KO method leads to asymptotically $L_2$-inconsistent calibration. This $L_2$-inconsistency can be remedied by modifying the original estimation procedure. A novel calibration method, called the $L_2$ calibration, is proposed and proven to be $L_2$-consistent and enjoys optimal convergence rate. A numerical example and some mathematical analysis are used to illustrate the source of the $L_2$-inconsistency problem.
\end{abstract}



\pagestyle{myheadings}
\thispagestyle{plain}
\markboth{Rui Tuo and C. F. Jeff Wu}{A theoretical framework for calibration}

\section{Introduction}
Because of the advances in complex mathematical models and fast computer codes, experiments on a computer, or referred to as computer experiments in the statistical literature, have become popular in engineering and scientific investigations. Computer simulations can be much faster or less costly than running physical experiments. Furthermore, physical experiments can be difficult to conduct as in the detonation of explosive materials or even infeasible when only rare events like land slide or hurricane are observed. Therefore computer simulations can be a stand-alone tool or combined with (typically smaller) data from physical experiments or field observations. There are many successful applications of computer experiments as reported in the literature. For a review of the general methodologies and examples, see the books by \cite{santner2003design}, and \cite{fang2005design}, and the November 2009 issue of \textit{Technometrics}, which was devoted to computer experiments.

In this paper we consider the situations in which both physical experiments/observations and computer simulations are conducted and some input variables in the computer code are either unknown or unmeasured in the physical experiment. We refer to them as \textit{calibration parameters}. From the responses in physical experiments alone, we cannot estimate the true values of the calibration parameters. We can run the computer codes by choosing selected values of the calibration parameters. From the combined data of the two sources, we can make inference about the parameters. That is, we can use the physical responses to \textit{calibrate} the computer model. Apart from the calibration parameters, \textit{control variables} are also involved as in standard computer experiments \cite{santner2003design}.

We use a spot welding example from \cite{bayarri2007framework} to illustrate the control variables and calibration parameters. In resistance spot welding, two sheets of metal are compressed by water-cooled copper electrodes under an applied load, $L$. A direct current of magnitude $C$ is supplied to the sheets by two electrodes to create localized heating at the interface (called ``faying surface'') between the two sheets. The heat produced by the current flow across the faying surface leads to melting, and, after cooling, a weld "nugget" is formed. The size of nugget is taken as the response because it gives a good measure of the strength of the weld. Here $L$ and $C$ are considered as control variables. The contact resistance at the faying surface is a calibration parameter because it cannot be measured in physical experiments but can be used as an input variable to a finite element code called ANSYS. 

In their pioneering work Kennedy and O'Hagan \cite{kennedy2001bayesian} proposed a model to link the two data sources by employing Gaussian process models, which are commonly used in the computer experiments literature. Since its publication, this approach has received a great deal of attention in the statistical literature. See \cite{ bayarri2007computer,bayarri2007framework,han2009simultaneous,higdon2008computer,higdon2004combining,joseph2009statistical,wang2009bayesian}, among others. It has seen a variety of applications, including hydrology, radiological protection, cylinder implosion, spot welding, micro-cutting and climate prediction, which were reported in the papers mentioned above and also in \cite{goldstein2004probabilistic} and \cite{murphy2007methodology}. In spite of its importance as a methodology and significant practical impact, there has been no theoretical research on its modeling and estimation strategies. The \textit{main purpose} of this paper is to provide a theoretical framework that facilitates the study of issues of parametrization, estimation and modeling in the Kennedy-O'Hagan formulation. For simplicity, we shall refer to Kennedy-O'Hagan as KO in the rest of the paper.

The paper is organized as follows. Some basic notation and terminology are given in Section \ref{sec:preliminaries}. A new theoretical framework for the calibration problem and its connection to function approximation via Gaussian process modeling are given in Section \ref{sec:calibration}. In particular, the lack of identifiability of the calibration parameters is discussed and a well-defined notion of calibration parameters is proposed by using the $L_2$ distance projection. The $L_2$-consistency is defined as a justification for calibration methods. The KO modeling strategy is discussed in Section \ref{Sec least native norm}. In order to provide a clean and workable mathematical analysis, we consider in Section \ref{Sec KO} some simplifications of their original formulation. One is to drop the prior, which should not affect the general conclusions of our work because the information in the prior becomes negligible as the data gets larger. Thus we shall refer to calibration based on this simplification as the \textit{KO calibration}.  A key result is Theorem \ref{Th theta1}, which states that the likelihood calibration is asymptotically $L_2$-inconsistent according to our definition of the true calibration parameters. A numerical example is given to show that the $L_2$-inconsistency can have a dramatic effect in small samples and some mathematical analysis is given to shed some light on why this happens. See Section \ref{Sec KO} and \ref{sec:numerical}. To rectify the $L_2$-inconsistency problem, a modification of the KO calibration is proposed in Section \ref{Sec ModifiedKO} by introducing a scale parameter into its correlation function. When the scale parameter converges to $+\infty$ at a certain rate, $L_2$-consistency is restored (see Theorem \ref{Th calibration}). The convergence rate of the original (unmodified) KO calibration is given in Theorem \ref{Th rateKO} of Section \ref{sec:rate}. To achieve both $L_2$-consistency and optimal convergence rate, we introduce a new method called least $L_2$ distance calibration in Section \ref{Sec L2deterministic} and prove such properties in Theorem \ref{Th L2} for the case of cheap code, i.e., when the computer code can be evaluated with no cost. Its extension to expensive code is given in Theorem \ref{Th expensive} of Section \ref{Sec Expensive}. The convergence rate is slower than that in Theorem \ref{Th L2} because, in expensive code, there is cost in evaluating the code and an unknown function $y^s$ associated with the code needs to be estimated from data. Concluding remarks are given in Section \ref{sec:discussion}. Technical details are given in two appendices. Throughout the paper, mathematical tools and results in native space \cite{wendland2005scattered} are extensively used.

\section{Preliminaries}\label{sec:preliminaries}

For a convex and compact region $\Omega\subset \mathbf{R}^d$, let $C(\Omega)$ be the set of continuous functions over $\Omega$. For a multiple index $\alpha=(\alpha_1,\ldots,\alpha_d)$, define $|\alpha|=\alpha_1+\ldots+\alpha_d$. Given $x=(x_1,\ldots,x_d)$ and function $f$, we denote the partial derivatives of $f$ by
\begin{eqnarray*}
D^\alpha f:=\frac{\partial^{|\alpha|}}{\partial x_1^{\alpha_1}\cdots\partial x_d^{\alpha_d}}f.
\end{eqnarray*}
For integer $k>0$, define $C^k(\Omega)=\{f\in C(\Omega):D^\alpha f\in C(\Omega) \text{ for } |\alpha|\leq k\}$. For a function $f$ over $\Omega$, define the $L_2$ norm as $\|f\|_{L_2(\Omega)}=(\int_\Omega f^2)^{1/2}$ and the Sobolev norm as
\begin{eqnarray}
\|f\|_{H^k(\Omega)}=\sqrt{\sum_{|\alpha|\leq k}\|D^\alpha f\|^2_{L_2(\Omega)}}. \label{sobolev}
\end{eqnarray}
The Sobolev space $H^k(\Omega)$ consists of functions with finite $H^k(\Omega)$ norm value. The definition of the Sobolev spaces can be extended to the case where $k$ is a real number. Such spaces are called the fractional Sobolev spaces and we refer to \cite{adams2003sobolev} for details.

Functional approximation methods play an important role in the estimation of the calibration parameters.
In this article, we consider the method of \textit{kernel interpolation} \cite{fasshauer2011positive}. This method provides a good functional approximation when the design $\mathcal{D}$ consists of \textit{scattered} points, i.e., the design points do not have any regular structure.

Suppose $y$ is a smooth function over $\Omega$ and $y(x_1),\ldots,y(x_n)$ are observed.
A kernel interpolator $\hat{y}$ is built as follows. First choose a symmetric positive definite function $\Phi(\cdot,\cdot)$ over $\Omega\times\Omega$. Two common choices for $\Phi$ are the squared exponential family (also referred to as the Gaussian correlation family), with
\begin{eqnarray}\label{Gaussian}
\Phi(s,t;\phi)=\exp\{-\phi \|s-t\|^2\}
\end{eqnarray}
and the Mat\'{e}rn family \cite{stein1999interpolation}, with
\begin{eqnarray}\label{matern}
\Phi(s,t;\nu,\phi)=\frac{1}{\Gamma(\nu)2^{\nu-1}}\left(2\sqrt{\nu}\phi \|s-t\|\right)^\nu K_\nu\left(2\sqrt{\nu}\phi\|s-t\|\right),
\end{eqnarray}
where $K_\nu$ is the modified Bessel function of the second kind.
Let $\mathbf{\Phi}=(\Phi(x_i,x_j))_{i j}$. Since the function $\Phi$ is positive definite, the matrix $\mathbf{\Phi}$ is also positive definite. Thus the linear system about $u=(u_1,\ldots,u_n)^\text{T}$
\begin{eqnarray}
Y=\mathbf{\Phi}u\label{linear}
\end{eqnarray}
has a unique solution $u=\mathbf{\Phi}^{-1}Y$, where $Y=(y(x_1),\ldots,y(x_n))^\text{T}$. For $x\in\Omega$, let
\begin{eqnarray}
\hat{y}(x)=\sum_{i=1}^n u_i \Phi(x,x_i).\label{interpolator}
\end{eqnarray}
It can be verified that $\hat{y}(x)$ indeed interpolates $(x_i,y(x_i))$'s.

We call the kernel $\Phi$ \textit{stationary} if $\Phi(x_1,x_2)$ depends only on the difference $x_1-x_2$.
Another special case of the kernel interpolation is the \textit{radial basis function interpolation} \cite{buhmann2003radial}, in which the kernel function $\Phi(x_1,x_2)$ depends only on the distance $\|x_1-x_2\|$ as in (\ref{Gaussian}) or (\ref{matern}). The choice of the kernel function is critical to the performance of the interpolation. Cross-validation is a common method for choosing a suitable kernel function, see \cite{santner2003design,Rasmussen2006gaussian}. 

In computer experiments, \textit{Gaussian process models} are widely used as surrogate models for unknown functions \cite{sacks1989design}.
There is a known relationship between the kernel interpolation and the Gaussian process prediction \cite{anjyo2011rbf}. Suppose $z(x)$ is a Gaussian process on $\Omega$ with mean 0 and covariance function $\Phi$. Then given $Z=(z(x_1),\ldots,z(x_n))^\text{T}$, the predictive mean of $z(x)$ for any $x$ is
\begin{eqnarray}
E[z(x)|Z]=\Phi(x,\mathbf{x})^\text{T}\mathbf{\Phi}^{-1}Z,\label{Gaussianprocess}
\end{eqnarray}
where $\Phi(x,\mathbf{x})=(\Phi(x,x_1),\ldots,\Phi(x,x_n))^\text{T}$. It can be seen that $\hat{y}$ in (\ref{interpolator}) has the same form as the predictive mean in (\ref{Gaussianprocess}). For details, see the book \cite{santner2003design}.

\section{Calibration Problem}\label{sec:calibration}

Suppose we have a physical system with a vector of control variables $x$ as its input.
Denote the input domain of $x$ by $\Omega$, which is a convex and compact subset of $\mathbf{R}^d$.
The response of this system given $x$ is denoted as $y^p(x)$.
We call the physical system \textit{deterministic} if $y^p(x)$ is a fixed value for each $x\in\Omega$, and \textit{stochastic} if $y^p(x)$ is random for some $x$.
To study the response surface, we conduct experiments on a selected set of points $\{x_1,\ldots,x_n\}$. The set $\mathcal{D}=\{x_1,\ldots,x_n\}$ is called the experimental design or \textit{design} for brevity.

We also have a computer code to simulate the physical system. The input of this computer code consists of two types of variables: the control variable $x$ and the \textit{calibration variable} $\theta$. The latter represents inherent attributes of the physical system, which cannot be controlled in the physical experiment. Denote the input domain for $\theta$ by $\Theta$, which is a compact subset of $\mathbf{R}^q$. The computer code gives a deterministic function of $x$ and $\theta$, denoted as $y^s(x,\theta)$.

Computer experiments are usually much less costly than the corresponding physical experiments. In an ideal situation, a computer run only takes a short time so that we can run the computer code as many times as we want. Mathematically speaking, we call a computer code \textit{cheap} if the functional form for $y^s$ is \textit{known}. However, computer runs may be time-consuming so that we can only evaluate $y^s$ on a set of design points. In this case, an estimate $\hat{y}^s(\cdot)$ based on the observed $y^s$ values (and the corresponding input values) is needed and we call the computer code \textit{expensive}.

In many cases, the true value of the calibration parameters cannot be measured physically. For instance, material properties like porosity and permeability are important computer inputs in computational material simulations, which cannot be measured directly in physical experiments. A standard approach to identify those parameters is to use physical responses to adjust the computer outputs. Use of the physical response and computer output to estimate the calibration parameters is referred to as \textit{calibration}.

\subsection{Kennedy-O'Hagan Method}\label{Sec least native norm}
Kennedy and O'Hagan \cite{kennedy2001bayesian} was the first to propose a Bayesian framework for the estimation of the calibration parameters.
The original version of the Kennedy-O'Hagan method works for stochastic systems with expensive computer codes.
Denote the physical response by $y^p(x_i)$, for $i=1,\ldots,n$.
Kennedy and O'Hagan \cite{kennedy2001bayesian} supposes that the physical response follows independent normal distribution. Specifically, they suggests that
\begin{eqnarray}
y^p(x_i)=\zeta(x_i)+e_i,
\end{eqnarray}
where $\zeta(x_i)=E y^p(x_i)$ and $e_i\operatorname*{\sim}\limits^{i.i.d.} N(0,\varsigma^2)$ with an unknown $\varsigma\geq 0$.
Kennedy and O'Hagan \cite{kennedy2001bayesian} denotes the ``true'' value of the calibration parameter by $\theta_0$ and
proposes the following model to link $\zeta(\cdot)$ and $y^s(\cdot,\theta_0)$
\begin{eqnarray}
\zeta(\cdot)=\rho y^s(\cdot,\theta_0)+\delta(\cdot),\label{KOdeterministic}
\end{eqnarray}
where $\rho$ is an unknown regression coefficient, $\delta(\cdot)$ is an unknown discrepancy function.
Kennedy and O'Hagan \cite{kennedy2001bayesian} claims that $\delta$ is a \textit{nonzero} function because the computer code is built based on certain assumptions or simplifications which do not match the reality exactly.
Thus $y^p$ and $y^s$ are related via the model:
\begin{eqnarray}
y^p(x)=\rho y^s(x,\theta_0)+\delta(x)+e.\label{KOmodel}
\end{eqnarray}
As is typically done in the literature of computer experiments, they assume that $y^s(\cdot,\cdot)$ and $\delta(\cdot)$ are independent realizations of Gaussian processes. The use of Gaussian process modeling in computer experiment problems can be traced back to \cite{sacks1989design}. In Gaussian process modeling, we usually choose the Gaussian or Mat\'{e}rn correlation functions (see (\ref{Gaussian}) and (\ref{matern})) and regard the parameters like $\phi$ and $\nu$ as \textit{unknown} models parameters. Then $\theta_0$ can be estimated from (\ref{KOmodel}) through a Bayesian approach.

\subsection{$L_2$ Distance Projection}\label{sec:projection}

The aim of this work is to establish a theoretical framework for the calibration problems from a frequentist point of view, i.e., we regard $\zeta(\cdot),y^s(\cdot,\cdot)$ in (\ref{KOdeterministic}) as deterministic functions. For simplicity, we rewrite (\ref{KOdeterministic}) as
\begin{eqnarray}
\zeta(\cdot)=y^s(\cdot,\theta_0)+\delta(\cdot).
\end{eqnarray}
This does not make much difference because we can regard the term $\rho y^s(x,\theta)$ as the computer output with calibration parameters $(\rho,\theta)$.

From now on, we suppose the physical system is \textit{deterministic}, i.e., $y^p(x_i)=\zeta(x_i)$ or equivalently $e_i=0$. Then under the framework of \cite{kennedy2001bayesian}, the calibration problem can be formulated as
\begin{eqnarray}
y^p(x)=y^s(x,\theta_0)+\delta(x),\label{lackindentifiability}
\end{eqnarray}
where $\theta_0$ is the ``true'' value of the calibration parameter and $\delta$ is the \textit{discrepancy} between $y^p$ and $y^s(\cdot,\theta_0)$.

From a frequentist perspective, $\theta_0$ in (\ref{lackindentifiability}) is \textit{unidentifiable}, because the pair $(\theta_0,\delta(\cdot))$ cannot be uniquely determined even if $y^p(\cdot)$ and $y^s(\cdot)$ are known.
This identifiability issue is discussed in \cite{bayarri2007computer,bayarri2007framework,han2009simultaneous} and other papers.

The main purpose of this section is to provide a rigorous theoretical study on the estimation of calibration parameters. Given the lack of identifiability for $\theta$, we need to find a well-defined parameter in order to study the estimation problem. A standard approach when the model parameters are unidentifiable is to redefine the ``true'' parameter value as one that minimizes the ``distance'' between the model and the observed data. First define
\begin{eqnarray}
\epsilon(x,\theta):=y^p(x)-y^s(x,\theta).\label{epsilon}
\end{eqnarray}
Here we adopt the $L_2$ norm in defining the distance. If another distance measure is chosen, the proposed mathematical framework can be pursued in a similar manner.

\begin{definition}\label{Def 1}
The \textit{$L_2$ distance projection} of $\theta$ is given by
\begin{eqnarray}
\theta^*=\operatorname*{argmin}\limits_{\theta\in\Theta}\|\epsilon(\cdot,\theta)\|_{L_2(\Omega)},\label{least distance}
\end{eqnarray}
where $\epsilon$ is defined in (\ref{epsilon}). For brevity, we will also refer to $\theta^*$ as the \textit{$L_2$ projection}.
\end{definition}

In Definition \ref{Def 1} we find a $\theta$ value that minimizes the $L_2$ discrepancy between the physical and computer observations, because the true value of $\theta$ is not estimable. The value $\theta^*$ given by (\ref{least distance}) minimizes the average predictive error given by the computer code. This is relevant and useful since our interest lies in the prediction of the physical response.
One justification for choosing the $L_2$ norm comes from the common use of the quadratic loss criterion.
Suppose we want to predict the physical response at a new point $x_0$ and $x_0$ is uniformly distributed over $\Omega$. Then the expected quadratic loss given $\theta$ is
\begin{eqnarray}
\int_\Omega (y^p(x)-y^s(x,\theta))^2 d x=\|\epsilon(\cdot,\theta)\|^2_{L_2(\Omega)}.\label{quandratic loss}
\end{eqnarray}
Thus the $\theta$ value minimizing $(\ref{quandratic loss})$ is the $L_2$ distance projection $\theta^*$.

The value of $\theta^*$ depends on the norm that is used to measure the discrepancy between the physical and the computer observations. If the $L_2$ norm is generalized to an $L_p$ norm or some weighted version, the results in the paper are not affected. Detailed discussion on this will be deferred to Section \ref{sec:discussion}.
In (\ref{least distance}) we implicitly assumes that the (global) minimizer of $\|\epsilon(\cdot,\theta)\|_{L_2(\Omega)}$ is unique. We believe that this is a reasonable assumption in many calibration problems. The definition of the $L_2$ projection also put the parameter estimation problem into an ``engineering validation'' framework. However, we still keep the wording ``calibration'' because it is the standard terminology since the foundation work by Kennedy and O'Hagan \cite{kennedy2001bayesian}. For a related discussion, we refer to \cite{chang2014model,joseph2014engineering}.

Since the functional forms for $y^p$ and $y^s$ are unknown, $\theta^*$ cannot be obtained by solving (\ref{least distance}). For the problems with cheap computer code, we know $y^s$ and the function values of $y^p$ over a set of design points, denoted as $y^p(\mathcal{D})$. For the problems with expensive computer code, we know only $y^p(\mathcal{D})$ and the function values of $y^s$ over the design points for the computer simulation, denoted as $y^s(\mathcal{G})$. Call $\hat{\theta}$ a (deterministic) estimator of $\theta^*$, if $\hat{\theta}$ depends only on $(\mathcal{D},y^p(\mathcal{D}),y^s)$ for cheap code or on $(\mathcal{D},\mathcal{G},y^p(\mathcal{D}),y^s(\mathcal{G}))$ for expensive code.
For fixed $y^p$ and $y^s$, let $\{\hat{\theta}_n\}$ be a sequence of estimates given by a sequence of designs (given by either $\{\mathcal{D}_n\}$ or $\{(\mathcal{D}_n,\mathcal{G}_n)\}$). Then $\hat{\theta}_n$ is said to be \textit{$L_2$-consistent} if $\hat{\theta}_n$ tends to $\theta^*$ as the designs become dense over $\Omega$ or $(\Omega,\Omega\times\Theta)$. The term ``consistent'' or ``consistency'' is a misnomer but we keep it here because of its statistical implication.

\section{Frequentist Properties of the Kennedy-O'Hagan Model}\label{sec:KO}

In this section we examine the frequentist properties of the calibration model by \cite{kennedy2001bayesian}. Our theoretical analysis shows that the method is $L_2$-inconsistent. We also construct some examples to show that the Kennedy-O'Hagan method may produce unreasonable answers.

\subsection{Simplified KO calibration}\label{Sec KO}

In order to conduct a rigorous mathematical analysis, we make the following simplifications to the Kennedy-O'Hagan method. We refer to this simplified version as the simplified KO calibration, or \textit{KO calibration} for brevity.
\begin{enumerate}[(i)]
\item
The computer code is cheap.
\item
The physical system is deterministic.
\item
Without loss of generality, we can assume $\rho=1$ (, otherwise the unknown $\rho$ can be regarded as a calibration parameter).
The discrepancy function $\delta$ is a realization of a Gaussian process with mean 0 and the covariance function $\sigma^2\Phi$, where $\sigma^2$ is an unknown parameter and the function $\Phi$ is \textit{known}.
\item
Maximum likelihood estimation (MLE) is used to estimate $(\theta,\sigma^2)$ instead of Bayesian analysis.
\end{enumerate}

The assumption (i) on cheap code will be relaxed for the $L_2$ calibration in Section \ref{Sec Expensive}. The assumption (ii) on deterministic physical experiment can be relaxed but will require a separate treatment. Further remarks are deferred to the end of Section \ref{sec:discussion}.
In assumption (iv), we only consider the likelihood portion of the Bayesian formulation in order to have a clean and workable mathematical analysis. As will be discussed in Section \ref{sec:discussion}, this simplification should not affect the general message one may draw regarding the original Bayesian formulation. Not to break the flow, further comments and justifications for the assumptions will be deferred to the concluding section.

Under these assumptions, the functions $\epsilon(x_i,\cdot)$ are known for $i=1,\ldots,n$.
Then the likelihood function given in \cite{kennedy2001bayesian} can be simplified and it can be shown that the log-likelihood function for $(\theta,\sigma^2)$ here is given by
\begin{eqnarray}
l(\theta,\sigma^2;Y)=-\frac{n}{2}\log \sigma^2-\frac{1}{2}\log |\mathbf{\Phi}|-\frac{1}{2\sigma^2}\epsilon(\mathbf{x},\theta)^\text{T}\mathbf{\Phi}^{-1}\epsilon(\mathbf{x},\theta),\label{KOlikelihood}
\end{eqnarray}
where $\epsilon(\mathbf{x},\theta)=(\epsilon(x_1,\theta),\ldots,\epsilon(x_n,\theta))^\text{T}$, $\mathbf{\Phi}=(\Phi(x_i,x_j))_{i j}$ and $|\mathbf{\Phi}|$ denotes the determinant of $|\mathbf{\Phi}|$. For details on the likelihood functions of Gaussian process models, we refer to \cite{santner2003design,Rasmussen2006gaussian}.

Our study will employ the reproducing kernel Hilbert spaces (also called the native spaces) as the mathematical tool \cite{wendland2005scattered}. Given a symmetric and positive definite function $\Phi$, define the linear space
\begin{eqnarray*}
F_\Phi(\Omega)=\left\{\sum_{i=1}^N \beta_i\Phi(\cdot,x_i):N\in\mathds{N},\beta_i\in\mathbf{R},x_i\in \Omega\right\}
\end{eqnarray*}
and equip this space with the bilinear form
\begin{eqnarray*}
\Big\langle\sum_{i=1}^N \beta_i\Phi(\cdot,x_i),\sum_{j=1}^M \gamma_j\Phi(\cdot,y_j)\Big\rangle_\Phi:=\sum_{i=1}^N\sum_{j=1}^M\beta_i\gamma_j\Phi(x_i,y_j).
\end{eqnarray*}
Define the native space $\mathcal{N}_\Phi(\Omega)$ as the \textit{closure} of $F_\Phi(\Omega)$ under the inner product $\langle\cdot,\cdot\rangle_\Phi$. The inner product of $\mathcal{N}_\Phi(\Omega)$, denoted as $\langle\cdot,\cdot\rangle_{\mathcal{N}_\Phi(\Omega)}$, is induced by $\langle\cdot,\cdot\rangle_\Phi$. Define the native norm as $\|f\|_{N_\Phi(\Omega)}=\sqrt{\langle f,f\rangle_{\mathcal{N}_\Phi(\Omega)}}$. Some required properties of reproducing kernel Hilbert spaces are given in Appendix \ref{App RKHS} under Propositions \ref{prop power function}-\ref{prop scale}. Their equations are numbered as (\ref{integral equation}) to (\ref{norm bound}).

Direct calculation shows that the maximum likelihood estimation (MLE) for $\theta$ is
\begin{eqnarray}
\hat{\theta}_{KO}=\operatorname*{argmin}_{\theta\in\Theta}\epsilon(\mathbf{x},\theta)^\text{T}\mathbf{\Phi}^{-1}\epsilon(\mathbf{x},\theta).\label{MLE}
\end{eqnarray}
For fixed $\theta$, let $\hat{\epsilon}(\cdot,\theta)$ be the kernel interpolator for $\epsilon(\cdot,\theta)$ given by (\ref{interpolator}), i.e.,
\begin{eqnarray}
\hat{\epsilon}(\cdot,\theta)=\Phi(\cdot,\mathbf{x})^\text{T}\mathbf{\Phi}^{-1}\epsilon(\mathbf{x},\theta).\label{epsilonhat}
\end{eqnarray}
From the definition of the native norm, we have
\begin{eqnarray*}
\|\hat{\epsilon}(\cdot,\theta)\|^2_{\mathcal{N}_{\Phi}(\Omega)}= \epsilon(\mathbf{x},\theta)^\text{T}\mathbf{\Phi}^{-1}\mathbf{\Phi}\mathbf{\Phi}^{-1}\epsilon(\mathbf{x},\theta)= \epsilon(\mathbf{x},\theta)^\text{T}\mathbf{\Phi}^{-1}\epsilon(\mathbf{x},\theta),
\end{eqnarray*}
which, together with (\ref{MLE}), gives
\begin{eqnarray}
\hat{\theta}_{KO}=\operatorname*{argmin}_{\theta\in\Theta}\|\hat{\epsilon}(\cdot,\theta)\|^2_{\mathcal{N}_{\Phi}(\Omega)}.\label{KO}
\end{eqnarray}

Now we study the asymptotic properties for the KO calibration model. To this end, we require the design points to become dense over $\Omega$. This property is measured by the fill distance.

\begin{definition}
For a design $\mathcal{D}\in\Omega^n$, define its \textit{fill distance} as
\begin{eqnarray}
h(\mathcal{D}):=\max_{x\in\Omega}\min_{x_i\in\mathcal{D}}\|x_i-x\|.
\end{eqnarray}
\end{definition}

We use $\hat{\theta}_{KO}(\mathcal{D})$ to denote the estimator $\hat{\theta}_{KO}$ under $\mathcal{D}$.
Theorem \ref{Th theta1} gives the limiting value of $\hat{\theta}_{KO}(\mathcal{D})$.

\begin{theorem}\label{Th theta1}
Suppose there exists $v_{\theta}\in L_2(\Omega)$, such that
\begin{eqnarray}
\epsilon(x,\theta)=\int_\Omega \Phi(x,t)v_{\theta}(t)d t\label{vexist}
\end{eqnarray}
for any $\theta\in\Theta$.
Moreover, suppose $\sup\limits_{\theta\in\Theta}\|v_{\theta}\|_{L_2}<+\infty$, $\Phi$ has continuous second order derivatives, and there exists a unique $\theta'\in\Theta$ such that
\begin{eqnarray}
\|\epsilon(\cdot,\theta')\|_{\mathcal{N}_\Phi(\Omega)}=\inf\limits_{\theta\in\Theta}\|\epsilon(\cdot,\theta)\|_{\mathcal{N}_\Phi(\Omega)}. \label{thetaprime}
\end{eqnarray}
Then $\hat{\theta}_{KO}(\mathcal{D}_n)\rightarrow\theta'$, provided that $h(\mathcal{D}_n)\rightarrow 0$ as $n\rightarrow\infty$.
\end{theorem}

The condition $\epsilon(x,\theta)=\int_\Omega \Phi(x,t)v_{\theta}(t)d t$ implies that $\epsilon(\cdot,\theta)$ lies in a subset of $\mathcal{N}_\Phi(\Omega)$. See (\ref{integral equation}) and (\ref{v}) in Appendix \ref{App RKHS} and \cite{wendland2005scattered} for further discussions.

\subsection{$L_2$ norm or native norm?}

Because in general $\theta'\neq\theta^*$, according to Definition \ref{Def 1}, the KO calibration is \textit{not} $L_2$-consistent. A bigger issue is whether $\theta^*$ is an appropriate definition for the calibration parameters. For example, suppose we adopt $\theta'$ in (\ref{thetaprime}) as the ``true'' calibration parameters. Then the KO calibration $\hat{\theta}_{KO}$ can be declared consistent according to Theorem \ref{Th theta1}. The question is whether $\theta'$ can be used as a legitimate definition for the calibration parameters.
Here we remind that Kennedy and O'Hagan \cite{kennedy2001bayesian} states the goal of calibration is ``... adjusting the unknown parameter until the outputs of the (computer) model fit the observed data''. However, we will give an example, backed by mathematical theory, to show that the result given by KO calibration may not agree with their original purpose.

In view of the convergence result in Theorem \ref{Th theta1}, we first study \textit{how different the limiting value $\theta'$ of the KO calibration is from $\theta^*$}. To address this question, we consider the difference between the two norms $\|\cdot\|_{L_2(\Omega)}$ and $\|\cdot\|_{\mathcal{N}_\Phi(\Omega)}$. This difference is related to the eigenvalues of the
integral operator defined as
\begin{eqnarray}
\kappa(f)=\int_\Omega \Phi(\cdot,x)f(x)d x,\label{kappa}
\end{eqnarray}
for $f\in L_2(\Omega)$. Denote the eigenvalues of $\kappa$ by $\lambda_1\geq\lambda_2\geq\cdots$. Let $f_i$ be the eigenfunction associated with $\lambda_i$ with $\|f_i\|_{L_2(\Omega)}=1$. Then it can be shown that
\begin{eqnarray}
\|f_i\|^2_{\mathcal{N}_\Phi(\Omega)}=\langle f_i,\lambda_i^{-1}f_i\rangle^2_{L_2(\Omega)}=\lambda^{-1}_i,\label{fi}
\end{eqnarray}
where the first equality follows from (\ref{v}) and the fact that $\kappa(\lambda_i^{-1}f_i)=f_i$. It is known in functional analysis that $\kappa$ is a compact operator and therefore $\lim_{k\rightarrow\infty}\lambda_k=0$ \cite{conway1990course}. Thus $(\ref{fi})$ yields that
$\|f_i\|^2_{\mathcal{N}_\Phi(\Omega)}/\|f_i\|^2_{L_2(\Omega)}=\lambda^{-1}_i\rightarrow \infty$
as $i\rightarrow\infty$. This leads to
\begin{eqnarray}
\sup_{f\in\mathcal{N}_\Phi(\Omega)}\frac{\|f\|_{\mathcal{N}_\Phi(\Omega)}}{\|f\|_{L_2(\Omega)}}=\infty,\label{unbounded}
\end{eqnarray}
which implies that there are functions $f$ with arbitrarily small $L_2$ norm while their $\mathcal{N}_\Phi$ norm is bounded away from zero. Therefore, by Definition \ref{Def 1}, the KO calibration can give results that are \textit{far from} the $L_2$ distance projection.
The following example shows that this effect can indeed be dramatic.

\begin{example}\label{ex eigen}
Consider a calibration problem with a three-level calibration parameter.
Let $\Omega=[-1,1]$, $\Phi(x_1,x_2)=\exp\{-(x_1-x_2)^2\}$. By using a numerical method, we obtain the eigenvalue and eigenfunction of $\kappa$ defined in $(\ref{kappa})$. The first and second eigenvalues are $\lambda_1=1.546$ and $\lambda_2=0.398$. For a better visual effect, we use the eigenfunctions whose $L_2$ norms are $\sqrt{20}$. We plot the first and second eigenfunctions of $\kappa$ in Figure \ref{Fig eigen}. We also plot $\sin 2\pi x$ for later comparison. Suppose we have three different computer codes. Denote the discrepancy between the physical response and each of the computer output by $\epsilon_1$, $\epsilon_2$ and $\epsilon_3$ respectively. Suppose $\epsilon_1,\epsilon_2,\epsilon_3$ are the three functions given in Figure \ref{Fig eigen}, i.e., $\epsilon_1$ and $\epsilon_2$ are the first and second eigenfunction of $\kappa$, and $\epsilon_3$ is $\sin 2\pi x$. Then which code is the best? From $\|\epsilon_1\|_{L_2(\Omega)}=\|\epsilon_2\|_{L_2(\Omega)}=\sqrt{20}$ and $\|\epsilon_3\|_{L_2(\Omega)}=1$, the third computer code is the best according to Definition \ref{Def 1}.

\begin{figure}[h]
\centering
\includegraphics[width=0.6 \textwidth]{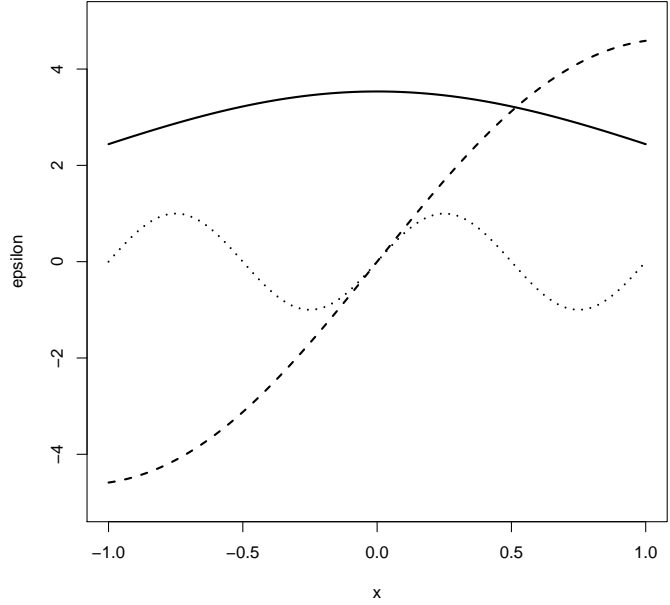}
\caption{Three functions considered in Example \ref{ex eigen}. The solid and dashed lines are the first and second eigenfunction of $\kappa$ with $L_2$ norm $\sqrt{20}$ respectively. The dotted line shows function $\sin 2\pi x$.}\label{Fig eigen}
\end{figure}

However, by using a Gaussian process model with the same correlation function $\Phi$, we get a different result.
By $(\ref{MLE})$, maximizing the likelihood function is equivalent to minimizing the pivoted sum of square (PSS): $\varepsilon^\text{T}_i\mathbf{\Phi}^{-1}\varepsilon_i$, where $\varepsilon_i=(\epsilon_i(x_1),\ldots,\epsilon_i(x_n))^\text{T}$ for $i=1,2,3$. We choose a space-filling design of 11 points, given by $x_j=-1+(j-1)/5$ for $j=1,\ldots,11$. The PSSs are $12.594$ for $i=1$, $57.908$ for $i=2$, and $17978.65$ for $i=3$. Thus the KO calibration will choose the first code because it has the smallest PSS value. This demonstrates that the likelihood-based method can give very different rankings of the competing codes from the $L_2$ projection. From Figure \ref{Fig eigen} we can see that $|\epsilon_3(x)|$ is smaller than $|\epsilon_1(x)|$ and $|\epsilon_2(x)|$ for every $x$, i.e., the point-wise prediction error for the third code is uniformly smaller than the first two. Therefore, the KO calibration made a wrong choice. This also gives a good justification for choosing the $L_2$ norm rather than the native norm in Definition \ref{Def 1}.
\end{example}

To give a more general explanation for the phenomenon in Example \ref{ex eigen}, we consider the situations where the Mat\'ern correlation functions defined by (\ref{matern}) are used. Corollary \ref{coro matern} in Appendix \ref{App RKHS} shows that for the Mat\'{e}rn correlation functions, the reproducing kernel Hilbert  space $\mathcal{N}_\Phi(\Omega)$ equals to the (fractional) Sobolev space $H^{\nu+d/2}(\Omega)$ and the two norms are equivalent for $\nu\geq 1$, where the Sobolev norm is defined by (\ref{sobolev}).
Note that the Sobolev norm can be big for a function with wild oscillation even when its $L_2$ norm is small (the same as that shown by (\ref{unbounded})). Thus, the Sobolev norm, which is equivalent to the native norm in this context, is not a good measure of discrepancy, because we only care about the magnitude of the discrepancy, not its oscillation. Therefore it is not suitable to use the KO calibration in most practical problems. For Gaussian correlation function, this problem is even more serious because the reproducing kernel Hilbert spaces generated by Gaussian kernels can be embedded into any Sobolev space (which can be shown by Proposition \ref{prop fourier}).

The phenomenon shown in Example \ref{ex eigen} can also be interpreted with the help of the Karhunen-Lo\`eve expansion of Gaussian processes.
Suppose $z(x)$ is a Gaussian process with mean 0 and covariance function $\Phi(\cdot,\cdot)$. Let $\lambda_i$ be the eigenvalues of integral operator $\kappa$ in (\ref{kappa}) with $\lambda_1\geq\lambda_2\geq\cdots$ and $f_i$ be the eigenfunction associated with $\lambda_i$ with $\|f_i\|_{L_2[-1,1]}=1$. Then the Karhunen-Lo\`eve theorem states that $z(\cdot)$ admits the follow expansion
\begin{eqnarray}
z(x)=\sum_{i=1}^\infty \lambda_i Z_i f_i(x),\label{KLexpansion}
\end{eqnarray}
where $Z_i$'s are independent and identically distributed standard normal random variables, and the convergence is in $L_2[-1,1]$. The expression (\ref{KLexpansion}) explains why $f_1$ is more likely to be a realization of $z(\cdot)$ among other functions with the same $L_2$ norm, i.e., $f_1$ yields the greatest likelihood value. To see this, we truncate (\ref{KLexpansion}) at a sufficiently large $i$, denoted as $K$, and obtain
\begin{eqnarray}
z(x)\approx \sum_{i=1}^K \lambda_i Z_i f_i(x).
\end{eqnarray}
Since $\{f_i\}$ forms an orthogonal basis in $L_2[-1,1]$, we can approximate any function in $L_2[-1,1]$ by
\begin{eqnarray*}
f(x)\approx \sum_{i=1}^K \langle f,f_i\rangle_{L_2[-1,1]} f_i(x).
\end{eqnarray*}
Thus the statement ``$f$ is a realization of $z(\cdot)$'' approximately yields $\lambda_i Z_i=\langle f,f_i\rangle_{L_2[-1,1]}$. This event has a probability density proportional to
\begin{eqnarray}
p(\lambda_i Z_i=\langle f,f_i\rangle_{L_2[-1,1]}):=\exp\left\{-\sum_{i=1}^K \langle f,f_i\rangle_{L_2[-1,1]}^2/ (2 \lambda_i^2)\right\},\label{likelihoodf}
\end{eqnarray}
because $Z_i\sim N(0,1)$.
Thus (\ref{likelihoodf}) can be regarded as a multiply of the probability density of sampling $f$ from $z(\cdot)$ approximately. Now we can find the function with the largest density value among a set of $f$ with the same $L_2$ norm, say $\|f\|_{L_2[-1,1]}=1$ without loss of generality. Let us assume $\lambda_1>\lambda_2$, which is true for the covariance function we discussed in Example \ref{ex eigen}. Because $1=\|f\|_{L_2[-1,1]}^2=\sum_{i=1}^\infty \langle f,f_i\rangle_{L_2[-1,1]}^2$, the function $f$ maximizes $(\ref{likelihoodf})$ should satisfy $\langle f,f_1\rangle_{L_2[-1,1]}=1$ and $\langle f,f_i\rangle_{L_2[-1,1]}=0$ for $i=2,3,\ldots$. Such a function is $f_1$.

\subsection{Numerical Study on Kennedy-O'Hagan Method with Estimated Correlation Function}\label{sec:numerical}
To give a more realistic comparison, we extend the study in Example \ref{ex eigen} by considering the frequentist version of the original Kennedy-O'Hagan method, in which the function $\Phi$ is estimated as well. Specifically, we suppose that $\Phi$ depends on a model parameter $\phi$, denoted as $\Phi_\phi(\cdot,\cdot)$. Then the log-likelihood function is
\begin{eqnarray}
\text{~~~~~~~}l(\theta,\sigma^2,\phi;Y)=-\frac{n}{2}\log \sigma^2-\frac{1}{2}\log |\mathbf{\Phi}_\phi|-\frac{1}{2\sigma^2}\epsilon(\mathbf{x},\theta)^\text{T}\mathbf{\Phi}_\phi^{-1}\epsilon(\mathbf{x},\theta),\label{philikelihood}
\end{eqnarray}
where $\mathbf{\Phi}_\phi=(\Phi_\phi(x_i,x_j))_{i j}$. By substituting the analytical form of $\hat{\sigma}^2$ into (\ref{philikelihood}), we obtain the log-likelihood function with respect to $(\theta,\phi)$:
\begin{eqnarray}
l(\theta,\phi;Y)=-\frac{n}{2}\log \epsilon(\mathbf{x},\theta)^\text{T}\mathbf{\Phi}_\phi^{-1}\epsilon(\mathbf{x},\theta)-\frac{1}{2}\log |\mathbf{\Phi}_\phi|.\label{phiMLE}
\end{eqnarray}
As in Section \ref{Sec KO}, we estimate $(\theta,\phi)$ by using the maximum likelihood.
In this subsection we present some numerical results, which show that even when $\Phi$ is estimated, the frequentist KO method still suffers from the problem demonstrated in Example \ref{ex eigen}.

\begin{figure}[h]
\centering
\includegraphics[width=0.6 \textwidth]{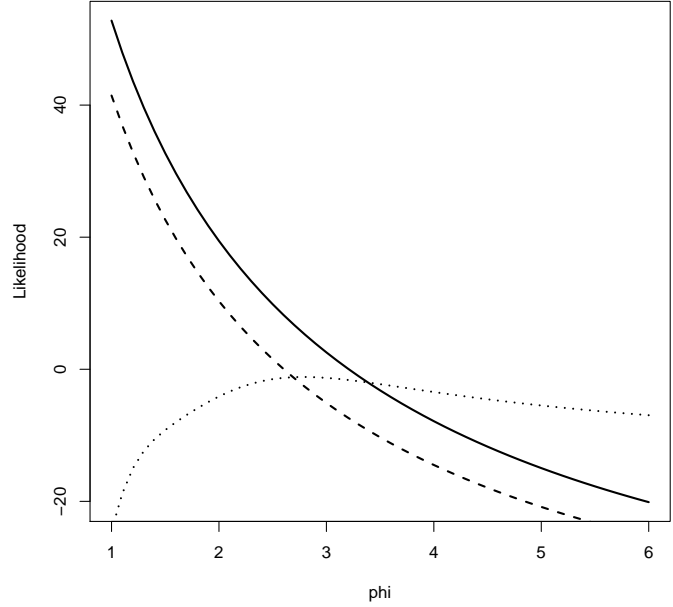}
\caption{Log-likelihood functions for the three functions $\epsilon_1,\epsilon_2,\epsilon_3$ given in Example \ref{ex eigen} are plotted in solid, dashed and dotted lines respectively.}\label{Fig likelihood}
\end{figure}

\addtocounter{example}{-1}
\begin{example}[continued]
We use the same true functions $\epsilon_1,\epsilon_2,\epsilon_3$ and design points as in Example \ref{ex eigen}. We compute the log-likelihood functions given in (\ref{phiMLE}) with $\Phi_\phi(x_1,x_2)=\exp\{-\phi(x_1-x_2)^2\}$. Denote the log-likelihood function in (\ref{phiMLE}) by $l(\theta,\phi)$, where $\theta=1,2,3$ correspond to the candidate functions $\epsilon_1,\epsilon_2,\epsilon_3$. The functions $l(\theta,\cdot)$ are plotted in Figure \ref{Fig likelihood} for $\theta=1,2,3$. From the figure we can see that $\sup_{\theta\in\{1,2,3\},\phi\in[1,6]}l(\theta,\phi)=l(1,1)$. Therefore the frequentist KO method will pick $\epsilon_1$, which gives the same (incorrect) result as in Example \ref{ex eigen}.
\end{example}

It can be seen from Figure (\ref{Fig likelihood}) that the log-likelihood functions  $l(1,\phi)$ and $l(2,\phi)$ are monotonic decreasing. This suggests that if $\phi$ ranges over $(0,6]$, the MLE of $\phi$ can be even smaller. Unfortunately, we are not able to calculate the likelihood value for a very small $\phi$ because the correlation matrix becomes nearly singular. Our current numerical experience show that likelihood value keeps growing as $\phi$ decreases. We conjecture that the likelihood function is unbounded in this case, although we are not able to prove it so far. From Figure (\ref{Fig likelihood}) it can also be seen that if we fix a relatively large $\phi$, say $\phi>4$, the likelihood values give a correct order of the $L_2$ discrepancy. This is not occasional. In Section \ref{Sec ModifiedKO} we will prove that such a modified version of the KO model leads to $L_2$-consistent estimation.

\section{Asymptotic Results: Cheap Code}\label{Sec deterministic}

Theorem \ref{Th theta1} is the first asymptotic result we present in this article. In this section we shall study other convergence properties, assuming that the computer code is cheap as in Section \ref{Sec KO}.

\subsection{Modified KO Calibration}\label{Sec ModifiedKO}

Given the wide-spread use of the Gaussian process modeling in calibration problems (as in the KO model), a fundamental question is whether we can modify it to rectify its $L_2$-inconsistency problem.
For convenience we assume a stationary Gaussian process model $Y(x)$.
The correlation of $Y$ is given by
\begin{eqnarray*}
R(x)=Corr(Y(\cdot+x),Y(\cdot)),
\end{eqnarray*}
where $R$ is a positive definite kernel over $\mathbf{R}^d$.
The \textit{Fourier transform} \cite{stein1971introduction} is a useful tool for studying stationary kernels.
We will use the notation $\mathcal{N}_R(\Omega)$ instead of $\mathcal{N}_{R(\cdot-\cdot)}(\Omega)$ for simplicity.

\begin{definition}\label{fourier}
For $f\in L_1(\mathbf{R}^d)$ define the Fourier transform by
\begin{eqnarray*}
\tilde{f}(w):=(2\pi)^{-d/2}\int_{\mathbf{R}^d} f(x)e^{-i w^\text{T}x}d x.
\end{eqnarray*}
\end{definition}

From Theorem \ref{Th theta1}, it can be seen that the KO calibration is not $L_2$-consistent if the correlation function $R$ is fixed. In order to construct $L_2$-consistent calibration, we should use a sequence of $R$ functions indexed by $n$, denoted by $R_n$.
From (\ref{v}), the $\|\cdot\|_{\mathcal{N}_\Phi(\Omega)}$ norm becomes $\|\cdot\|_{L_2}$ only when $\Phi(x_1,x_2)=\delta(x_1-x_2)$, where $\delta$ denotes the Dirac delta function. We need the convergence $R_n(x)\rightarrow\delta(x)$ in order to obtain $L_2$-consistency. An easy way to achieve this convergence is to introduce a scale parameter.
Suppose $R(\cdot;\phi)$ is a family of correlation functions on $\mathbf{R}^d$ with $\phi>0$. Call $\phi$ a \textit{scale parameter} if $R(x;\phi)=R(\phi x;1)$ for any $\phi>0$ and any $x\in\mathbf{R}^d$. Most correlation families like Gaussian or Mat\'{e}rn family satisfy these conditions.

Write $R_n(x)=R(x;\phi_n)$. Let $\hat{\theta}(R_n,\mathcal{D}_n)$ be the estimate of $\theta$ given by the KO calibration using correlation function $R_n$ under design $\mathcal{D}_n$, referred to as the \textit{modified KO calibration}.
The $L_2$-consistency requires $\phi_n\rightarrow\infty$. But to ensure the convergence of the interpolation, $\phi_n$ cannot diverge too fast.
The next theorem suggests that the modified KO calibration is $L_2$-consistent if we choose a suitable increasing rate for $\phi_n$. Define the convolution $f*f(x)=\int_{\mathbf{R}^d}f(x-t)f(t)d t$ for any $f\in L_2(\mathbf{R}^d)$. We list the required regularity conditions before stating the theorem.

\begin{enumerate}[{A}1{:}]
\item $\sup_{x\in \Omega,\theta\in\Theta}\|\nabla_x \epsilon(x,\theta)\|<+\infty$, where $\nabla_x \epsilon(x,\theta)$ is the gradient of $\epsilon(x,\theta)$ with respect to $x$.
\item There exists $\phi_0>0$ such that $\sup_{\theta\in\Theta}\|\epsilon(\cdot,\theta)\|_{\mathcal{N}_{R(\cdot;\phi_0)*R(\cdot;\phi_0)}(\Omega)}<+\infty$.
\item $R(\cdot;1)$ is integrable and $\sup_{w\neq 0,\alpha\geq 1} \tilde{R}(\alpha w)/\tilde{R}(w)<+\infty$, where $\tilde{R}$ is the fourier transform of $R(\cdot;1)$. 
\end{enumerate}

\begin{theorem}\label{Th calibration}
Suppose conditions (A1-A3) are satisfied and
$R(\cdot;1)$ has continuous derivatives. Then $\hat{\theta}(R_n,\mathcal{D}_n)\rightarrow \theta^*$ if $\phi_n\rightarrow +\infty$ and $\phi_n h(\mathcal{D}_n)\rightarrow 0$.
\end{theorem}

Although the modified KO calibration is $L_2$-consistent, its implementation relies on a prespecified sequence $\phi_n$. For a calibration problem with a fixed sample size, there is no theoretical guidelines on choosing the $\phi$ value. A more practical procedure is given in Section \ref{Sec L2deterministic}, namely, a novel calibration method that is $L_2$-consistent and does not rely on the choice of the kernel function.

\subsection{Convergence Rate}\label{sec:rate}

Under the assumption of Theorem \ref{Th theta1}, we can employ (\ref{improved}) in Proposition \ref{prop power function} to show that the interpolation error given by a kernel $\Phi$ with $2k$ derivatives is equivalent to $O(h^{2 k}(\mathcal{D}_n))$. Given this rate, Theorem \ref{Th rateKO} shows that the KO calibration, which converges to $\theta'$ in Theorem \ref{Th theta1}, reaches the same rate. Let $\theta=(\theta_1,\ldots,\theta_q)^\text{T}$.

\begin{theorem}\label{Th rateKO}
Under the conditions of Theorem \ref{Th theta1}, suppose $\Phi$ has $2 k$ continuous derivatives. We assume that $\theta'$ is an interior point of $\Theta$ and there exist a neighborhood $U\subset \Theta$ of $\theta'$, and functions $D_i v_\theta,D_{i j} v_\theta\in C(\Omega)$ such that
\begin{eqnarray}
\frac{\partial \epsilon}{\partial \theta_i}(x,\theta)=\int_\Omega \Phi(x,t) D_i v_{\theta}(t)d t, \label{v1}\\
\frac{\partial^2 \epsilon}{\partial \theta_i\partial \theta_j}(x,\theta)=\int_\Omega \Phi(x,t) D_{i j}v_{\theta}(t)d t,\label{v2}
\end{eqnarray}
for $x\in\Omega,\theta\in U$ and $1\leq i,j\leq q$. Moreover,
\begin{eqnarray}
&&\sup_{\theta\in U,1\leq i,j\leq q}\{\|D_i v_\theta\|_{L_2(\Omega)},\|D_{i j} v_\theta\|_{L_2(\Omega)}\}<\infty, \text{ and}\label{supv}\\
&&\frac{\partial^2}{\partial \theta\partial\theta^\text{T}}\|\epsilon(\cdot,\theta')\|^2_{\mathcal{N}_\Phi(\Omega)} \text{ is invertible}. \label{calibrationinvertible}
\end{eqnarray}
Then $\|\hat{\theta}_{KO}(\mathcal{D}_n)-\theta'\|=O(h^{2 k}(\mathcal{D}_n))$.
\end{theorem}

The conditions (\ref{v1}) and \ref{v2} enhance the condition (\ref{vexist}) in Theorem \ref{Th theta1} by assuming the differentiability of $\epsilon$ and the interchangeability of a derivative and an integral (differentiation under the integral sign; i.e.,Leibniz integral rule).

Noting that consistency is a necessary requirement for an estimator, we would like to find an estimator that is consistent and attains the same convergence rate as in Theorem \ref{Th rateKO}. In the following subsection we find an estimator that guarantees $L_2$-consistency and full efficiency.

\subsection{Least $L_2$ Distance Calibration}\label{Sec L2deterministic}

Let $\hat{y}^p$ be the kernel interpolator defined in $(\ref{interpolator})$ for $y^p$ under design $\mathcal{D}$.
Define the \textit{least $L_2$ distance calibration} by
\begin{eqnarray}
\hat{\theta}_{L_2}(\mathcal{D})=\operatorname*{argmin}_{\theta\in\Theta}\|\hat{y}^p(\cdot)-y^s(\cdot,\theta)\|_{L_2(\Omega)}.\label{L2calibration}
\end{eqnarray}
For brevity, we will also refer to it as the \textit{$L_2$ calibration}. \cite{han2009simultaneous} used the $L_2$ norm in a different context, i.e., choosing an optimal tuning parameter value to minimize the $L_2$ norm of the observed discrepancy $\hat{y}^p-y^s$.
Theorem \ref{Th L2} shows that $\hat{\theta}_{L_2}(\mathcal{D}_n)$ converges to the $L_2$ projection $\theta^*$ at the optimal rate.

\begin{theorem}\label{Th L2}
Suppose $\theta^*$ is the unique solution to (\ref{least distance}) and an interior point of $\Theta$; $y^p\in\mathcal{N}_\Phi(\Omega)$; $\frac{\partial^2}{\partial \theta\partial\theta^\text{T}}\|\epsilon(\cdot,\theta^*)\|^2_{L_2(\Omega)}$ is invertible; $\Phi$ has $2 k$ continuous derivatives; and there exists a neighborhood $U\subset \Theta$ of $\theta^*$ such that $y^s\in L_2(\Omega\times U)$ and $y^s(x,\cdot)\in C^2(U)$ for $x\in\Omega$.
Then as $h(\mathcal{D}_n)\rightarrow 0$,
\begin{eqnarray}
\|\hat{\theta}_{L_2}(\mathcal{D}_n)-\theta^*\|=O(h^k(\mathcal{D}_n)).\label{L2origin}
\end{eqnarray}
Furthermore, if there exists $v\in L_2(\Omega)$ such that $y^p(x)=\int_\Omega\Phi(x,t)v(t) d t$ for all $x\in\Omega$, then the convergence rate can be improved to
\begin{eqnarray}
\|\hat{\theta}_{L_2}(\mathcal{D}_n)-\theta^*\|=O(h^{2 k}(\mathcal{D}_n)).\label{L2improved}
\end{eqnarray}
\end{theorem}

By comparing the results and conditions in Theorems \ref{Th rateKO} and \ref{Th L2}, we can make the following observations. First, $\|\hat{\theta}_{KO}(\mathcal{D}_n)-\theta'\|$ and $\|\hat{\theta}_{L_2}(\mathcal{D}_n)-\theta^*\|$ enjoy the convergence rate $O(h^{2k}(\mathcal{D}_n))$ under similar conditions. Second, the $L_2$ calibration has the additional property that, even under much less restrictive conditions, it still enjoys convergence, though at a slower rate. But this slower rate is optimal under these conditions because the interpolator $\hat{y}^p$ has the same convergence rate given by (\ref{ordinary}).

\section{Least $L_2$ Distance Calibration for Expensive Code}\label{Sec Expensive}

Now we turn to the case of expensive computer code for which $y^s$ cannot be evaluated for infinitely many times. In this situation we need another surrogate model for $y^s$.
Note that the input space for a computer run is $\Omega\times \Theta\subset\mathbf{R}^{d+q}$. Let $\mathcal{G}$ be the set of design points for the computer experiment with its fill distance $h(\mathcal{G})$. Suppose $\Theta$ is convex.
Choose a positive definite function $\Psi$ over $(\Omega\times\Theta)\times(\Omega\times\Theta)$. For kernel $\Psi$ and design $\mathcal{G}$, let $\hat{y}^s$ be the interpolate for $y^s$ defined by (\ref{interpolator}).
Then we can define the $L_2$ calibration in a similar way:
\begin{eqnarray}
\hat{\theta}_{L_2}(\mathcal{D},\mathcal{G})&:=&\operatorname*{argmin}_{\theta\in\Theta}\|\hat{y}^p(\cdot)-\hat{y}^s(\cdot,\theta)\|_{L_2(\Omega)}. \label{L2calibrationexpensive}
\end{eqnarray}
Note that the only difference from the definition in (\ref{L2calibration}) is the replacement of $y^s$ by its interpolate $\hat{y}^s$ in (\ref{L2calibrationexpensive}).

We want to study the asymptotic behavior of the $L_2$ calibration for expensive computer code.
First we need to extend the definition of $\theta^*$ in (\ref{least distance}) to
\begin{eqnarray*}
\theta^*(\mathcal{G})&:=&\operatorname*{argmin}_{\theta\in\Theta}\|y^p(\cdot)-\hat{y}^s(\cdot,\theta)\|_{L_2(\Omega)}.\label{L2G}
\end{eqnarray*}
Then we have
\begin{eqnarray}
\|\hat{\theta}_{L_2}(\mathcal{D},\mathcal{G})-\theta^*\|\leq \|\hat{\theta}_{L_2}(\mathcal{D},\mathcal{G})-\theta^*(\mathcal{G})\|+ \|\theta^*(\mathcal{G})-\theta^*\|.\label{triangle}
\end{eqnarray}
If we regard the interpolate $\hat{y}^s$ as the true computer output, $\theta^*(\mathcal{G})$ can be viewed as an ``$L_2$ projection''.
Following similar steps as in the proof of Theorem \ref{Th L2}, we can prove that
\begin{eqnarray}
\|\hat{\theta}_{L_2}(\mathcal{D}_n,\mathcal{G}_n)-\theta^*(\mathcal{G}_n)\|=O(h^k(\mathcal{D}_n))\label{expensiveterm1}
\end{eqnarray}
under the same conditions in Theorem \ref{Th L2}.
It remains to find a bound for $\|\theta^*(\mathcal{G})-\theta^*\|$.
The following theorem shows that its rate is slower than that in Theorem \ref{Th L2}.

\begin{theorem}\label{Th expensive}
Suppose $\theta^*$ is the unique solution to (\ref{least distance}); $\Psi$ has $2k'$ continuous derivatives with $k'\geq 3$; $\theta^*$ is an interior point of $\Theta$; $y^p\in L_2(\Omega)$; $y^s\in\mathcal{N}_{\Psi}(\Omega\times\Theta)$ and $\frac{\partial^2}{\partial \theta\partial\theta^\text{T}}\|\epsilon(\cdot,\theta^*)\|^2_{L_2(\Omega)}$ is invertible. Then
$\|\theta^*(\mathcal{G}_n)-\theta^*\|=O(h^{k'-1}(\mathcal{G}_n))$ as $h(\mathcal{G}_n)\rightarrow 0$.
\end{theorem}

Theorem \ref{Th expensive}, together with (\ref{triangle})-(\ref{expensiveterm1}), yields the following result on the convergence rate of the $L_2$ calibration for expensive computer code.

\begin{theorem}
Under the assumptions of Theorems \ref{Th L2} and \ref{Th expensive},
\begin{eqnarray*}
\|\hat{\theta}_{L_2}(\mathcal{D}_n,\mathcal{G}_n)-\theta^*\|=O(\max(h^k(\mathcal{D}_n),h^{k'-1}(\mathcal{G}_n))).
\end{eqnarray*}
\end{theorem}

\section{Further Discussions and Remarks}\label{sec:discussion}

This paper provides the first theoretical framework for studying modeling and estimation of the calibration parameters in statistical models that are motivated by and closely related to the original Kennedy-O'Hagan \cite{kennedy2001bayesian} approach. Being the first piece of such work and because of the space limitations, it has left some issues to be further considered. The definition of $L_2$-consistency in this paper is based on using $\theta^*$ in Definition \ref{Def 1} as the ``true'' calibration parameters. The same mathematical results should hold if a different positive definite metric is employed in defining $\theta^*$. The technical details may be more involved but the same lines of arguments can be used to obtain similar results.

One may also consider other definitions of $\theta^*$ by replacing the $L_2$ norm by alternatives such as the $L_p$ norms. The least $L_p$ norm calibration can be defined by straightforward modifications of (\ref{L2calibration}) and (\ref{L2calibrationexpensive}) and its efficiency can be proved following similar arguments to those of Theorem \ref{Th L2} and \ref{Th expensive}. There also exist norms which have different physical meanings. For example, if the oscillation of the predictor is of interest, a Sobolev norm, which may relate to the energy of the system, would be appropriate. The efficient estimation in such a framework would require a separate investigation.

In spite of the $L_2$-inconsistency calibration results, the Kennedy-O'Hagan method (which gives a calibration estimator converging to $\theta'$) can give a good prediction for $y^p(\cdot)$. This is supported by the upper bound in Proposition \ref{prop power function}. For a function with a smaller $N_\Phi(\Omega)$ norm, the kernel interpolate is likely to provide a better approximation because the upper bound given in (\ref{ordinary}) is smaller. This implies that it is easier to approximate the function $\epsilon(\cdot,\theta')$ than $\epsilon(\cdot,\theta^*)$. For example, the solid and dashed curves in Figure \ref{Fig eigen} can be more easily estimated than the (more fluctuating) dotted curve, although the former two have greater point-wise absolute values.

Next we turn to the discussions on the simplifications (i), (iii) and (iv) of assumptions made in Section \ref{Sec KO}. We have relaxed (i) in Section \ref{Sec Expensive} for the $L_2$ calibration. What about a similar extension to the KO calibration for the expensive code? As the situation becomes more messy (i.e., the need of estimating the function $y^s$), there are reasons to believe that the procedure will remain $L_2$-inconsistent but the mathematical details can be more daunting. However, until further analysis is done, we are not sure if the KO calibration would converge to $\theta'$. Regarding (iii), the first part on assuming $\rho =1$ can be easily relaxed to an unknown $\rho$ because $\rho$ can be included as part of the calibration parameters in the theoretical analysis. Its second part on assuming the $\Phi$ function is known can also be relaxed but a rigorous analysis will require more work. If we assume that the parameters for the $\Phi$ function (such as those in (\ref{Gaussian})-(\ref{matern})) lie in a compact set outside zero, our analysis should still be applicable because the sequence of estimated values of these parameters should have a nonzero  limit point. For a related discussion, see \cite{bull2011convergence}. Regarding assumption (iv), we can deal with the original Bayesian formulation in the KO paper by bringing back the prior information. Some heuristic calculations suggest that the prior for $\theta$ should have no effect on the asymptotic results given in Theorem \ref{Th theta1}, provided that $\theta'$  lies on the support of the prior distribution. Because of the space limitations, such extensions are deferred to future work.

In theorem \ref{Th rateKO}, \ref{Th L2}, \ref{Th expensive}, convergence rates are proven to be the fill distance to the power of some quantity related to the differentiability of the kernel function. This implies that, for infinitely differentiable kernel functions, such as the Gaussian covariance kernel (\ref{Gaussian}), the rate of convergence can be faster than any power function of the fill distance. In fact, by applying the error estimate for Gaussian kernels in \cite{wendland2005scattered}, a parallel development can show that an exponential rate of convergence can be achieved by using a Gaussian kernel.

Our work can be extended in another direction. We assume in (\ref{lackindentifiability}) that the physical system is a deterministic  function $y^p (x)$ for each $x$. To what extent can the present work be extended to a stochastic physical system, where $y^p (x)$ is  random for some or all $x$? This extension can be found in \cite{tuo2014efficient}. The new work is a major endeavor because there are three major differences. First, while the current work deals with deterministic functions, the stochastic work deals with random functions. As a result, the required mathematical tools are quite different: the current work employs techniques in native spaces \cite{wendland2005scattered}, while the latter employs the techniques of weak convergence. Finally, the statistical methods and results are different: the current work shows the efficiency of the interpolation-based $L_2$ calibration, while the latter proposes a novel method based on smoothing splines in the reproducing kernel Hilbert spaces and shows its semiparametic efficiency. Because of these major differences, the present work cannot be viewed as
a special case of the work in \cite{tuo2014efficient}.
In the situation where noisy presents, it seems that we are going to lose Theorem \ref{Th theta1} (i.e., the MLE of the KO model may not converge), see Example 1 in \cite{tuo2014efficient}. In other words, to prove the asymptotic theory for the KO model, deterministic physical experiments is a necessary assumption. Although it is more realistic to assume that the physical observations are noisy, the current paper gives a motivation for studying these new calibration methods rather than the Kennedy-O'Hagan's approach in view of the latter's $L_2$-inconsistency property.

A method alterative to the $L_2$ calibration proceeds by minimizing $\sum_{i=1}^n(y^p_i-\hat{y}^s(x_i,\theta))^2$ directly. We refer to this method as the ordinary least squares (OLS) method. Under the present context, it can be shown that the OLS method is $L_2$-consistent if $\{x_i\}$ is a mutually independent random sequence uniformly distributed over $\Omega$ or a quasi-Monte Carlo sequence. However, the rate of convergence stated in Theorem \ref{Th L2} cannot be attained by OLS. For the physical experiments with measurement error, a modified version of the $L_2$ calibration is also more efficient than the OLS method, see \cite{tuo2014efficient}. Note the in computer experiment and uncertainty quantification, a fast convergence is highly appreciated because the experiment is expensive in general \cite{xiu2010numerical}. This provides some justification of using the $L_2$ calibration over the OLS method.

\appendix
\section{Reproducing Kernel Hilbert Spaces}\label{App RKHS}

We give a brief summary of properties regarding reproducing kernel Hilbert space (RKHS). See \cite{wendland2005scattered} for details. First, if $\Omega$ is compact and there exists $v\in L^2(\Omega)$, such that
\begin{eqnarray}
f(x)=\int_\Omega \Phi(x,t)v(t)d t,\label{integral equation}
\end{eqnarray}
then $f\in\mathcal{N}_\Phi(\Omega)$ and for any $g\in \mathcal{N}_\Phi(\Omega)$,
\begin{eqnarray}
\langle f,g\rangle_{\mathcal{N}_\Phi(\Omega)}=\int_\Omega v(x)g(x)d x.\label{v}
\end{eqnarray}

The existence of $v$ can be guaranteed if $f\in\mathcal{N}_{\Phi*\Phi}(\Omega)$ \cite{haaland2012accurate}, where $\Phi*\Phi(x)=\int_{\mathbf{R}^d}\Phi(x-t)\Phi(t)d t$ is the convolution. Furthermore, they show that in this situation there exists a continuous function $v$ satisfying $(\ref{integral equation})$ and
\begin{eqnarray}
\|v\|_{L_2(\Omega)}\leq \|f\|_{\mathcal{N}_{\Phi*\Phi}(\Omega)}.\label{L2}
\end{eqnarray}

Wendland \cite{wendland2005scattered} discusses the error estimates of the kernel interpolation.
First, the following equality
\begin{eqnarray}
\|\hat{y}\|^2_{\mathcal{N}_\Phi(\Omega)}+\|y-\hat{y}\|^2_{\mathcal{N}_\Phi(\Omega)}=\|y\|^2_{\mathcal{N}_\Phi(\Omega)},\label{norm inequality}
\end{eqnarray}
follows from the projective property of RKHS.
Proposition \ref{prop power function} gives the error estimates for the interpolation and the native norm. As before, let $h(\mathcal{D})$ be the fill distance of the design points.

\begin{proposition}[Wendland \cite{wendland2005scattered}, p. 181]\label{prop power function}
Suppose that $\Phi$ has $2 k$ continuous derivatives. Then there exists a constant $C_\Phi$ such that
\begin{eqnarray}
\sup\limits_{x\in\Omega}|D^\alpha y(x)-D^\alpha \hat{y}(x)|\leq C_\Phi h^{k-|\alpha|}(\mathcal{D})\|y\|_{\mathcal{N}_\Phi(\Omega)},\label{ordinary}
\end{eqnarray}
if $y\in\mathcal{N}_\Phi$ and $|\alpha|\leq k$, where $\hat{y}$ is defined by $(\ref{interpolator})$; $C_\Phi$ is independent of $X$ and $y$; and $x_i$ is any component of $x$. Furthermore, if there exists $v\in L_2(\Omega)$, such that $y(x)=\int_\Omega \Phi(x,t)v(t)d t$.
Then the error bounds can be improved as follows:
\begin{eqnarray}
\sup\limits_{x\in\Omega}|y(x)-\hat{y}(x)|\leq C_\Phi h^{2 k}(\mathcal{D})\|y\|_{\mathcal{N}_\Phi(\Omega)},\label{improved}\\
\|y-\hat{y}\|_{\mathcal{N}_\Phi(\Omega)}\leq C_\Phi h^k(\mathcal{D})\|v\|_{L_2(\Omega)}.\label{norm bound}
\end{eqnarray}
\end{proposition}

We now turn to the extension or restriction of native spaces to another region. Assume that we have two convex regions $\Omega_1\subset \Omega_2\subset\mathbf{R}^d$ and $\Phi$ is a positive definite kernel over $\Omega_2\times\Omega_2$.

\begin{proposition}[Wendland \cite{wendland2005scattered}, p. 169]\label{prop extension}
Each function $f\in \mathcal{N}_\Phi(\Omega_1)$ has a natural extension to a function $Ef\in\mathcal{N}_\Phi(\Omega_2)$. Furthermore, $\|Ef\|_{\mathcal{N}_\Phi(\Omega_2)}=\|f\|_{\mathcal{N}_\Phi(\Omega_1)}$.
\end{proposition}

\begin{proposition}[Wendland \cite{wendland2005scattered}, p. 170]\label{prop restriction}
The restriction $f|\Omega_1$ of any function $f\in\mathcal{N}_\Phi(\Omega_2)$ is contained in $\mathcal{N}(\Omega_1)$ with $\|f|\Omega_1\|_{\mathcal{N}_\Phi(\Omega_1)}\leq\|f\|_{\mathcal{N}_\Phi(\Omega_2)}$.
\end{proposition}

Usually we assume the kernel function has the form $\Phi(x,y)=R(x-y)$, where $R$ is continuous and integrable over $\mathbf{R}^d$.
Denote the Fourier transform of $R$ by $\tilde{R}$. Since $R$ is symmetric, $\tilde{R}$ is real and $R$ can be recovered from $\tilde{R}$. Proposition \ref{prop fourier} shows that the RKHS $\mathcal{N}_R(\mathbf{R}^d)$ can be represented by using Fourier transforms.

\begin{proposition}[Wendland \cite{wendland2005scattered}, p. 139]\label{prop fourier}
Suppose that $R\in C(\mathbf{R}^d)\cap L_1(\mathbf{R}^d)$ is a real-valued positive definite function. Then
\begin{eqnarray*}
\mathcal{N}_R(\mathbf{R}^d)=\{f\in L_2(\mathbf{R}^d)\cap C(\mathbf{R}^d):\tilde{f}/\sqrt{\tilde{R}}\in L_2(\mathbf{R}^d)\},
\end{eqnarray*}
with the inner product given by
\begin{eqnarray*}
\langle f,g\rangle_{\mathcal{N}_R(\mathbf{R}^d)}=(2\pi)^{-d/2}\int_{\mathbf{R}^d}\frac{\tilde{f}(w)\overline{\tilde{g}(w)}}{\tilde{R}(w)}d w.
\end{eqnarray*}
\end{proposition}

Under certain conditions, the RKHSs are related to the (fractional) Sobolev spaces. Let $[a]$ denote the integer part of a real number $a$.

\begin{proposition}[Wendland \cite{wendland2005scattered}, p. 201]\label{prop sobolev}
Suppose there exist constants $c_1,c_2$ and $\tau$, such that the kernel $R$ satisfies
\begin{eqnarray*}
c_1(1+\|w\|^2)^{-\tau}\leq \tilde{R}(w)\leq c_2(1+\|w\|^2)^{-\tau},
\end{eqnarray*}
for $w\in\mathbf{R}$ with $[\tau]>d/2$. Then $\mathcal{N}_R(\Omega)=H^\tau(\Omega)$ with equivalent norms.
\end{proposition}

Now consider the Mat\'ern correlation family $R_{\nu,\phi}$ given by (\ref{matern}). Applying Theorem 6.13 (p. 76) of \cite{wendland2005scattered}, after some direct calculations we find that the Fourier transformation of this family is
\begin{eqnarray*}
\tilde{R}_{\nu,\phi}(w)=2^{d/2}(4\nu\phi^2)^\nu\frac{\Gamma(\nu+d/2)}{\Gamma(\nu)}(4\nu\phi^2+\|w\|^2)^{-(\nu+d/2)}.
\end{eqnarray*}
Thus as a consequence of Proposition \ref{prop sobolev}, we obtain the following corollary.

\begin{corollary}\label{coro matern}
For $[\nu+d/2]>d/2$, the RKHS generated by the Mat\'ern correlation function (\ref{matern}) equals the Sobolev space $H^{\nu+d/2}(\Omega)$ with equivalent norms.
\end{corollary}

Let $R_\theta(x)=R(\theta x)$ for $\theta>0$. The next result, given by \cite{haaland2012accurate}, shows that, under certain conditions, (\ref{norm bound}) can be expressed in a more direct manner.

\begin{proposition}\label{prop scale}
Suppose $y\in\mathcal{N}_{R_\theta*R_\theta}(\Omega)$, and $\hat{y}$ is the kernel interpolator given by $R_\theta$. If $R$ has $k$ continuous derivatives, then
\begin{eqnarray*}
\|y-\hat{y}\|_{\mathcal{N}_{R_\theta}(\Omega)}\leq C_R \theta^{k/2}h^{k/2}(\mathcal{D})\|y\|_{\mathcal{N}_{R_\theta*R_\theta}(\Omega)},
\end{eqnarray*}
where $C_R$ is independent of $X$, $y$ and $\theta$.
\end{proposition}

\section{Technical Proofs}\label{App proof}

In this section we provide the formal proofs for Theorem \ref{Th theta1}, \ref{Th calibration}, \ref{Th rateKO}, \ref{Th L2}, \ref{Th expensive}.

\subsection{Proof of Theorem \ref{Th theta1}}
Let $\hat{\epsilon}_n$ be the kernel interpolator for $\epsilon$ under design $\mathcal{D}_n$.
From (\ref{KO}) and (\ref{thetaprime}), it suffices to prove that $\|\hat{\epsilon}_n(\cdot,\theta)\|_{\mathcal{N}_\Phi(\Omega)}^2$ converges to $\|\epsilon(\cdot,\theta)\|_{\mathcal{N}_\Phi(\Omega)}^2$ uniformly with respect to $\theta\in\Theta$. From Proposition \ref{prop power function}, we have
\begin{eqnarray}
\|\hat{\epsilon}_n(\cdot,\theta)\|_{\mathcal{N}_\Phi(\Omega)}^2-\|\epsilon(\cdot,\theta)\|_{\mathcal{N}_\Phi(\Omega)}^2 &=&\|\hat{\epsilon}_n(\cdot,\theta)-\epsilon(\cdot,\theta)\|^2_{\mathcal{N}_\Phi(\Omega)}\nonumber\\
&\leq& C_\Phi^2 h^2(\mathcal{D}_n)\|v_\theta\|^2_{L_2(\Omega)}\nonumber\\
&\leq& C_\Phi^2 h^2(\mathcal{D}_n)\sup\limits_{\theta\in\Theta}\|v_\theta\|^2_{L_2(\Omega)},\label{beta}
\end{eqnarray}
where the first equality follows from the identity $\langle\epsilon(\cdot,\theta),\hat{\epsilon}_n(\cdot,\theta)-\epsilon(\cdot,\theta)\rangle_{\mathcal{N}_\Phi(\Omega)}=0$.
The right side of $(\ref{beta})$ goes to $0$ as $n\rightarrow\infty$ and is independent of $\theta$. This completes the proof.
\hfill $\square$

\subsection{Proof of Theorem \ref{Th calibration}}
Without loss of generality, we assume $\phi_0=1$.
Let $R_\phi(x)=R(x;\phi)$ and $Q_\phi=\big(\phi^d R(\cdot;\phi))*(\phi^d R(\cdot;\phi)\big)$. The Fourier transform of $Q_\phi$ is
\begin{eqnarray}
\widetilde{Q_\phi}=(2\pi)^{d/2}\phi^{2 d}\widetilde{R_\phi}^2.\label{convolution}
\end{eqnarray}
We first study the relationship between $\|\cdot\|_{\mathcal{N}_{Q_1}(\Omega)}$ and $\|\cdot\|_{\mathcal{N}_{Q_\phi}(\Omega)}$, for $\phi>1$.
For any $f\in\mathcal{N}_{Q_1}(\Omega)$, by Proposition \ref{prop extension}, there exists an extension $Ef\in\mathcal{N}_{Q_1}(\mathbf{R}^d)$, such that $\|f\|_{\mathcal{N}_{Q_1}(\Omega)}=\|Ef\|_{\mathcal{N}_{Q_1}(\mathbf{R}^d)}$.
Then we have
\begin{eqnarray}
\|f\|^2_{\mathcal{N}_{Q_\phi}(\Omega)}
&\leq&\|Ef\|^2_{\mathcal{N}_{Q_\phi}(\mathbf{R}^d)}\nonumber\\
&=&(2 \pi)^{-d/2}\int_{\mathbf{R}^d}\frac{|\widetilde{Ef}(w)|^2}{\widetilde{Q_\phi}(w)}d w \nonumber\\
&=&(2 \pi)^{-d/2}\int_{\mathbf{R}^d}\frac{|\widetilde{Ef}(w)|^2}{(2\pi)^{d/2}\phi^{2 d}\widetilde{R_\phi}^2(w)}d w \nonumber\\
&=&(2 \pi)^{-d/2} \int_{\mathbf{R}^d}\frac{|\widetilde{Ef}(w)|^2}{(2\pi)^{d/2}\tilde{R}^2(w/\phi)}d w\nonumber\\
&\leq& (2 \pi)^{-d/2} \sup_{w\neq 0}\{\tilde{R}(\phi w)/\tilde{R}(w)\}^2 \int_{\mathbf{R}^d}\frac{|\widetilde{Ef}(w)|^2}{(2\pi)^{d/2}\tilde{R}^2(w)}d w\nonumber\\
&=&(2 \pi)^{-d/2} \sup_{w\neq 0}\{\tilde{R}(\phi w)/\tilde{R}(w)\}^2 \int_{\mathbf{R}^d}\frac{|\widetilde{Ef}(w)|^2}{\widetilde{Q_1}(w)}d w\nonumber\\
&=&\sup_{w\neq 0}\{\tilde{R}(\phi w)/\tilde{R}(w)\}^2\|Ef\|^2_{\mathcal{N}_{Q_1}(\mathbf{R}^d)}\nonumber\\
&\leq&\sup_{w\neq 0,\alpha\leq 1}\{\tilde{R}(\alpha w)/\tilde{R}(w)\}^2\|f\|^2_{\mathcal{N}_{Q_1}(\Omega)},\label{convolution ineq}
\end{eqnarray}
where the first inequality follows from Proposition \ref{prop restriction}; the first equality follows from Proposition \ref{prop fourier}; the second equality follows from (\ref{convolution}); the third equality follows from the fact that $\tilde{R_\phi}(w)=\phi^{-d}\tilde{R}(w/\phi)$; the second inequality follows from factoring out $\{\tilde{R}(\phi w)/\tilde{R}(w)\}^2$; the fourth equality follows from $(\ref{convolution})$; the fifth equality follows from Proposition \ref{prop fourier}; and the last inequality follows from condition A3 and Proposition \ref{prop extension}.

 Let $C_0=\int_{\mathbf{R}^d}R(x;1)d x$. Define integral operator $\kappa_\phi: L_2(\Omega)\mapsto L_2(\Omega)$ by
\begin{eqnarray*}
\kappa_\phi(f)(x)=C_0^{-1}\phi^{d}\int_{\Omega}R(x-y;\phi)f(y)d y, & x\in \Omega,
\end{eqnarray*}
for any $\phi>0$.
Obviously $\kappa_\phi$ is a self-adjoint operator, i.e., $\langle f, \kappa_\phi(g)\rangle=\langle \kappa_\phi(f), g\rangle$ for any $f,g\in L_2(\Omega)$.
For any $x\in \Omega$, let $x-\Omega=\{x-x_0:x_0\in \Omega\}$ and $\phi(x-\Omega)=\{\phi(x-x_0):x_0\in \Omega\}$. First we show that for any interior point $x$ in $\Omega$,
\begin{eqnarray}
&&\lim_{\phi\rightarrow+\infty}C^{-1}_0\phi^d\int_{\Omega} R(x-y;\phi)d y\nonumber\\
&=&\lim_{\phi\rightarrow+\infty}C^{-1}_0\phi^d\int_{\mathbf{R}^d}I(y\in x-\Omega)R(y;\phi)d y\nonumber\\
&=&\lim_{\phi\rightarrow+\infty}C^{-1}_0\phi^d\int_{\mathbf{R}^d}I(y\in x-\Omega)R(\phi y;1) d y\nonumber\\
&=&\lim_{\phi\rightarrow+\infty}C^{-1}_0\int_{\mathbf{R}^d}I(y\in\phi(x-\Omega))R(y;1) d y\nonumber\\
&=&C^{-1}_0 \int_{\mathbf{R}^d}R(y;1)d y=1,\label{density}
\end{eqnarray}
where the fourth equality follows from the dominated convergence theorem and the fact that $x-\Omega$ contains a neighborhood of $0$.

Let $r_n=C^{-1}_0\phi^d\int_{\Omega} R(x-y;\phi)d y-1$. Then for any interior point $x$ in $\Omega$,
\begin{eqnarray}
&&\sup_{\theta}|\kappa_\phi(\epsilon(\cdot,\theta))(x)-\epsilon(x,\theta)|\nonumber\\
&=&\sup_{\theta}\Big|C^{-1}_0\phi^d\int_{\Omega} R(x-y;\phi)\epsilon(y,\theta)d y\nonumber
\\&&-C^{-1}_0\phi^d\int_{\Omega} R(x-y;\phi)\epsilon(x,\theta)d y\nonumber+r_n \epsilon(x;\theta)\Big|\nonumber\\
&\leq&\sup_{\theta}\Big|C^{-1}_0\int_{\mathbf{R}^d}I(y\in\phi(x-\Omega))R(y;1)\nonumber\\&&\{\epsilon(x-y/\phi,\theta)-\epsilon(x,\theta)\} d y\Big|+\sup_{\theta}|r_n \epsilon(x,\theta)|\nonumber\\
&\leq&\phi^{-1}\sup_{\theta}\|\nabla_x\epsilon(x,\theta)\|\Big|\int_{\mathbf{R}^d}I(y\in\phi(x-\Omega))R(y;1)\|y\| d y\Big|\label{kf uniform}\\&&+|r_n|\sup_{\theta}|\epsilon(x,\theta)|\nonumber,
\end{eqnarray}
where the last inequality follows from the mean value theorem and condition A1. Using the dominated convergence theorem, we have $\int_{\mathbf{R}^d}I(y\in\phi(x-\Omega))R(y;1)(\|y\|/\phi) d y\rightarrow 0$, since $y/\phi$ lies in the bounded set $ x-\Omega$. This shows that the first term in (\ref{kf uniform}) tends to 0. The second term in (\ref{kf uniform}) also tends to 0 because of (\ref{density}). Thus $(\ref{kf uniform})$ shows that $\kappa_\phi(\epsilon(\cdot,\theta))(x)$ converges to $\epsilon(x,\theta)$ uniformly with respect to $\theta$.

From the definition of native spaces, it is easily seen that
\begin{eqnarray}
\|f\|^2_{\mathcal{N}_{K}(\Omega)}=c\|f\|^2_{\mathcal{N}_{cK}(\Omega)}\label{K scale}
\end{eqnarray}
 for any $f$, $K$ and $c>0$.
 Since condition A2 holds, by applying $(\ref{convolution ineq})$ to $\epsilon(\cdot,\theta)$, we see that for any $\phi>1$ and $\theta\in\Theta$, $\epsilon(\cdot,\theta)\in \mathcal{N}_{Q_\phi}(\Omega)$.
Consequently, from $(\ref{integral equation})$-$(\ref{L2})$, for any $\theta\in\Theta$ and $\phi>1$, there exists $v_{\phi,\theta}\in L_2(\Omega)$, such that $\epsilon(x,\theta)=\int_\Omega R(x-t;\phi)v_{\phi,\theta}(t) d t$.
Thus
\begin{eqnarray}
\epsilon(x,\theta)&=&\int_\Omega (C_0^{-1}\phi^d)R(x-t;1)(C_0\phi^{-d})v_{\phi,\theta}(t) d t\nonumber\\&=&\kappa((C_0\phi^{-d})v_{\phi,\theta})(x). \label{epsilon and v}
\end{eqnarray}
Applying $(\ref{L2})$ and $(\ref{convolution ineq})$, we have
\begin{eqnarray}
\|C_0\phi^{-d}v_{\phi,\theta}\|_{L_2(\Omega)}\leq\|\epsilon(\cdot,\theta)\|_{\mathcal{N}_{Q_\phi}(\Omega)}\nonumber\\
\leq\sup_{w\neq 0,\alpha\leq 1}\{\tilde{R}(\alpha w)/\tilde{R}(w)\}^2\|\epsilon(\cdot,\theta)\|^2_{\mathcal{N}_{Q_1}(\Omega)}<+\infty.\label{v bound}
 \end{eqnarray}
Then
\begin{eqnarray}
&&C_0\phi^{-d}\|\epsilon(\cdot,\theta)\|^2_{\mathcal{N}_{R_\phi}(\Omega)}-\|\epsilon(\cdot,\theta)\|^2_{L_2 (\Omega)}\nonumber\\
&=&\Big\langle\epsilon(\cdot,\theta),C_0\phi^{-d}v_{\phi,\theta}\Big\rangle_{L_2 (\Omega)}- \Big\langle\epsilon(\cdot,\theta),\epsilon(\cdot,\theta)\Big\rangle_{L_2 (\Omega)}\nonumber\\
&=&\Big\langle\epsilon(\cdot,\theta),C_0\phi^{-d}v_{\phi,\theta}-\epsilon(\cdot,\theta)\Big\rangle_{L_2 (\Omega)}\nonumber\\
&=&\Big\langle\kappa_\phi(C_0\phi^{-d}v_{\phi,\theta}),C_0\phi^{-d}v_{\phi,\theta}-\epsilon(\cdot,\theta)\Big\rangle_{L_2 (\Omega)}\nonumber\\
&=&\Big\langle C_0\phi^{-d}v_{\phi,\theta},\kappa_\phi\big(C_0\phi^{-d}v_{\phi,\theta}-\epsilon(\cdot,\theta)\big)\Big\rangle_{L_2 (\Omega)}\nonumber\\
&=&\Big\langle C_0\phi^{-d}v_{\phi,\theta},\epsilon(\cdot,\theta)-\kappa_\phi\big(\epsilon(\cdot,\theta)\big)\Big\rangle_{L_2 (\Omega)}\nonumber\\
&\leq&\|C_0\phi^{-d}v_{\phi,\theta}\|_{L_2 (\Omega)}\|\epsilon(\cdot,\theta)-\kappa_\phi\big(\epsilon(\cdot,\theta)\big)\|_{L_2 (\Omega)}\nonumber\\
&\leq&\|C_0\phi^{-d}v_{\phi,\theta}\|_{L_2 (\Omega)} \Big\|\sup_{\theta\in\Theta}\big|\epsilon(\cdot,\theta)-\kappa_\phi\big(\epsilon(\cdot,\theta)\big)\big|\Big\|_{L_2 (\Omega)},
\end{eqnarray}
where the first equality follows from $(\ref{v})$; the third equality follows from the definition of $\kappa_\phi$; the fourth equality follows from the self-adjoint property of $\kappa_\phi$; the fifth equality follows from $(\ref{epsilon and v})$; the first inequality follows from Schwarz's inequality.
Note that $(\ref{v bound})$ gives the uniform upper bound of $\|C_0\phi^{-d}v_{\phi,\theta}\|_{L_2(\Omega)}$ with respect to $\theta$. Using the dominated convergence theorem and $(\ref{kf uniform})$, $\|\sup_{\theta\in\Theta}|\epsilon(\cdot,\theta)-\kappa_\phi(\epsilon(\cdot,\theta)\big)|\|_{L_2(\Omega)}\rightarrow 0$, as $\phi\rightarrow\infty$.

Now return to the settings of Theorem \ref{Th calibration}. Since $\phi_n\rightarrow +\infty$ as $n\rightarrow\infty$, applying $(\ref{K scale})$, we have
\begin{eqnarray}
\sup_{\theta\in\Theta}\Big|\|\epsilon(\cdot,\theta)\|^2_{\mathcal{N}_{C_0^{-1}\phi_n^{d}R_{\phi_n}}(\Omega)} -\|\epsilon(\cdot,\theta)\|^2_{L_2 (\Omega)}\Big|\rightarrow 0,\label{convergence 1}
\end{eqnarray}
as $n\rightarrow\infty$.
On the other hand, we have
\begin{eqnarray}
&&\sup_{\theta\in\Theta}\Big|\|\epsilon(\cdot,\theta)\|^2_{\mathcal{N}_{C_0^{-1}\phi^d_n R_{\phi_n}}(\Omega)}- \|\hat{\epsilon}_n(\cdot,\theta)\|^2_{\mathcal{N}_{C_0^{-1}\phi^d_n R_{\phi_n}}(\Omega)}\Big|\nonumber\\
&=&\sup_{\theta\in\Theta}\Big|\|\epsilon(\cdot,\theta)-\hat{\epsilon}_n(\cdot,\theta)\|^2_{\mathcal{N}_{C_0^{-1}\phi^d_n R_{\phi_n}}(\Omega)}\Big|\nonumber\\
&\leq& C_R^2 \sqrt{\phi_n h(\mathcal{D}_n)}\sup_{\theta\in\Theta}\|\epsilon(\cdot,\theta)\|^2_{\mathcal{N}_{C_0^{-2}Q_{\phi_n}}(\Omega)}\label{factor effect}\\
&\leq&\Big(C_R^2 C_0^2\sup_{w\neq 0,\alpha\leq 1}\{\tilde{R}(\alpha w)/\tilde{R}(w)\}^2 \sup_{\theta\in\Theta}\|\epsilon(\cdot,\theta)\|^2_{\mathcal{N}_{Q_1}(\Omega)}\Big)\sqrt{\phi_n h(\mathcal{D}_n)}\nonumber\\
&\rightarrow& 0,\label{convergence 2}
\end{eqnarray}
where the equality follows from (\ref{norm inequality}); the first inequality follows from Proposition \ref{prop scale}; the second inequality follows from (\ref{convolution ineq}); and the limiting relationship follows from conditions A2, A3 and the fact that $\phi_n h(\mathcal{D}_n)\rightarrow 0$. One may notice that (\ref{factor effect}) does not follows immediately from Proposition \ref{prop scale} because in Proposition \ref{prop scale} $\hat{y}$ is built using the kernel $R_\theta$ but here $\hat{\epsilon}_n(\cdot,\theta)$ is built using $R_{\phi_n}$ instead of $C^{-1}_0 \phi_n^d R_{\phi_n}$. However, this is not a serious problem because using $(\ref{linear})$ and $(\ref{interpolator})$, and after some simple calculations, we can verify that the two interpolators given by $R_{\phi_n}$ and $C^{-1}_0 \phi_n^d R_{\phi_n}$ respectively are equal to each other. Based on this equivalence, Proposition \ref{prop scale} is still applicable.

Now we can bound $\left|C_0\phi^{-d}_n\|\hat{\epsilon}_n(\cdot,\theta)\|^2_{\mathcal{N}_{R_{\phi_n}}(\Omega)}-\|\epsilon(\cdot,\theta)\|^2_{L_2 (\Omega)}\right|$ with
\begin{eqnarray*}
&&\sup_{\theta\in\Theta}\Big|C_0\phi^{-d}_n\|\hat{\epsilon}_n(\cdot,\theta)\|^2_{\mathcal{N}_{R_{\phi_n}}(\Omega)}-\|\epsilon(\cdot,\theta)\|^2_{L_2 (\Omega)}\Big|\\
&=&\sup_{\theta\in\Theta}\Big|\|\hat{\epsilon}_n(\cdot,\theta)\|^2_{\mathcal{N}_{C_0\phi^{-d}_n R_{\phi_n}}(\Omega)}-\|\epsilon(\cdot,\theta)\|^2_{L_2 (\Omega)}\Big|\\
&\leq&\sup_{\theta\in\Theta}\Big|\|\epsilon(\cdot,\theta)\|^2_{\mathcal{N}_{C_0^{-1}\phi^d_n R_{\phi_n}}(\Omega)}- \|\hat{\epsilon}_n(\cdot,\theta)\|^2_{\mathcal{N}_{C_0^{-1}\phi^d_n R_{\phi_n}}(\Omega)}\Big|\\&&+\sup_{\theta\in\Theta}\Big|\|\epsilon(\cdot,\theta)\|^2_{\mathcal{N}_{C_0^{-1}\phi_n^{d}R_{\phi_n}}(\Omega)} -\|\epsilon(\cdot,\theta)\|^2_{L_2 (\Omega)}\Big|\rightarrow 0,
\end{eqnarray*}
where the first equality follows from (\ref{K scale}); the inequality follows from the triangle inequality; and the limiting relationship follows from (\ref{convergence 1}) and (\ref{convergence 2}).
Therefore, we have established the following result
 \begin{eqnarray*}
 \hat{\theta}(R_n,\mathcal{D}_n)=\operatorname*{argmin}_{\theta\in\Theta}\|\hat{\epsilon}_n(\cdot,\theta)\|^2_{\mathcal{N}_{R_{\phi_n}}(\Omega)}\rightarrow \operatorname*{argmin}_{\theta\in\Theta}\|\epsilon(\cdot,\theta)\|^2_{L_2 (\Omega)}=\theta^*,
 \end{eqnarray*}
 as $n\rightarrow\infty$.
This completes the proof.\hfill $\square$

\subsection{Proof of Theorem \ref{Th rateKO}}
For the proof we need some inequalities. The first one follows immediately from (\ref{norm bound}) in Proposition \ref{prop power function}:
\begin{eqnarray}
\|\hat{\epsilon}_n(\cdot,\theta)-\epsilon(\cdot,\theta)\|_{\mathcal{N}_\Phi(\Omega)}\leq C_\Phi h^k(\mathcal{D}_n) \|v_\theta\|_{L_2(\Omega)}.\label{ineq1}
\end{eqnarray}
From the definition of $\hat{\epsilon}_n$ in (\ref{epsilonhat}), we have $\frac{\partial \hat{\epsilon}}{\partial \theta_j}(\cdot,\theta)=\Phi(\cdot,\mathbf{x})^\text{T}\mathbf{\Phi}^{-1}\frac{\partial \epsilon}{\partial \theta_j}(\mathbf{x},\theta)$. Because $\frac{\partial \hat{\epsilon}_n}{\partial \theta_j}(\cdot,\theta)$ is also spanned by the functions $\{\Phi(\cdot,x_i)\}$,
$\frac{\partial \hat{\epsilon}_n}{\partial \theta_j}(\cdot,\theta)$ is equal to the kernel interpolator for the pairs $(x_i,\frac{\partial \epsilon}{\partial \theta}(x_i,\theta))$. As a result of (\ref{norm bound}) in Proposition \ref{prop power function}, we obtain
\begin{eqnarray}
\left\|\frac{\partial(\hat{\epsilon}_n-\epsilon)}{\partial \theta_i}(\cdot,\theta)\right\|_{\mathcal{N}_\Phi(\Omega)}\leq C_\Phi h^k(\mathcal{D}_n) \|D_i v_\theta\|_{L_2(\Omega)}.\label{ineq2}
\end{eqnarray}
Similarly we have
\begin{eqnarray}
\left\|\frac{\partial^2(\hat{\epsilon}_n-\epsilon)}{\partial \theta_i\partial \theta_j}(\cdot,\theta)\right\|_{\mathcal{N}_\Phi(\Omega)}\leq C_\Phi h^k(\mathcal{D}_n) \|D_{i j} v_\theta\|_{L_2(\Omega)}.\label{ineq3}
\end{eqnarray}
As $\|\hat{\epsilon}_n(\cdot,\theta)\|^2_{\mathcal{N}_\Phi(\Omega)}$ is minimized at $\hat{\theta}_{KO}(\mathcal{D}_n)$, the Taylor expansion of $\frac{\partial}{\partial \theta}\|\hat{\epsilon}_n(\cdot,\theta)\|^2_{\mathcal{N}_\Phi(\Omega)}$ with respect to $\theta$ gives
\begin{eqnarray*}
0&=&\frac{\partial}{\partial \theta}\|\hat{\epsilon}_n(\cdot,\hat{\theta}_n)\|^2_{\mathcal{N}_\Phi(\Omega)}\nonumber\\
&=&\frac{\partial}{\partial \theta}\|\hat{\epsilon}_n(\cdot,\theta')\|^2_{\mathcal{N}_\Phi(\Omega)}+\bigg(\frac{\partial^2}{\partial \theta\partial \theta^\text{T}}\|\hat{\epsilon}_n(\cdot,\tilde{\theta}_n)\|^2_{\mathcal{N}_\Phi(\Omega)}\bigg)(\hat{\theta}_{KO}(\mathcal{D}_n)-\theta'),
\end{eqnarray*}
where $\tilde{\theta}_n$ is located between $\theta'$ and $\hat{\theta}_{KO}(\mathcal{D}_n)$. By Theorem \ref{Th theta1}, $\tilde{\theta}_n\rightarrow\theta'$. Thus
\begin{eqnarray}
&&\hat{\theta}_{KO}(\mathcal{D}_n)-\theta'\nonumber\\&=&-\bigg(\frac{\partial^2}{\partial \theta\partial\theta^\text{T}}\|\hat{\epsilon}_n(\cdot,\tilde{\theta}_n)\|^2_{\mathcal{N}_\Phi(\Omega)}\bigg)^{-1} \frac{\partial}{\partial \theta}\|\hat{\epsilon}_n(\cdot,\theta')\|^2_{\mathcal{N}_\Phi(\Omega)},\label{error}
\end{eqnarray}
where $\frac{\partial^2}{\partial \theta\partial\theta^\text{T}}\|\hat{\epsilon}_n(\cdot,\tilde{\theta}_n)\|^2_{\mathcal{N}_\Phi(\Omega)}$ is invertible because of assumption (\ref{calibrationinvertible}) and the fact that $\tilde{\theta}_n\rightarrow\theta'$.
Furthermore,
\begin{eqnarray}
&&\frac{\partial^2}{\partial \theta_i\partial \theta_j}\|\hat{\epsilon}_n(\cdot,\tilde{\theta}_n)\|^2_{\mathcal{N}_\Phi(\Omega)}-\frac{\partial^2}{\partial \theta_i\partial \theta_j}\|\epsilon(\cdot,\tilde{\theta}_n)\|^2_{\mathcal{N}_\Phi(\Omega)}\nonumber\\
&=&\frac{\partial^2}{\partial \theta_i\partial \theta_j}\|\hat{\epsilon}_n(\cdot,\tilde{\theta}_n)-\epsilon(\cdot,\tilde{\theta}_n)\|^2_{\mathcal{N}_\Phi(\Omega)}\nonumber\\
&=&2\Bigg\{\left\langle\frac{\partial^2 (\hat{\epsilon}_n-\epsilon)}{\partial \theta_i\partial \theta_j}(\cdot,\tilde{\theta}_n),(\hat{\epsilon}_n-\epsilon)(\cdot,\tilde{\theta}_n)\right\rangle\nonumber\\
&&+\left\langle\frac{\partial(\hat{\epsilon}_n-\epsilon)}{\partial \theta_i}(\cdot,\tilde{\theta}_n),\frac{\partial(\hat{\epsilon}_n-\epsilon)}{\partial \theta_j}(\cdot,\tilde{\theta}_n)\right\rangle\Bigg\}\nonumber\\
&\leq&2\Bigg\{\|\hat{\epsilon}_n(\cdot,\tilde{\theta}_n)-\epsilon(\cdot,\tilde{\theta}_n)\|_{\mathcal{N}_\Phi(\Omega)} \left\|\frac{\partial^2(\hat{\epsilon}_n-\epsilon)}{\partial \theta_i\partial \theta_j}(\cdot,\tilde{\theta}_n)\right\|_{\mathcal{N}_\Phi(\Omega)}\nonumber\\
&&+\left\|\frac{\partial(\hat{\epsilon}_n-\epsilon)}{\partial \theta_i}(\cdot,\tilde{\theta}_n)\right\|_{\mathcal{N}_\Phi(\Omega)} \left\|\frac{\partial(\hat{\epsilon}_n-\epsilon)}{\partial \theta_j}(\cdot,\tilde{\theta}_n)\right\|_{\mathcal{N}_\Phi(\Omega)} \Bigg\},\label{tends0}
\end{eqnarray}
where the first equality follows from (\ref{norm inequality}). Invoking (\ref{ineq1})-(\ref{ineq3}), and the condition $h(\mathcal{D}_n)\rightarrow 0$ in Theorem \ref{Th theta1}, (\ref{tends0}) tends to 0. This results in
\begin{eqnarray}
\frac{\partial^2}{\partial \theta\partial\theta^\text{T}}\|\hat{\epsilon}_n(\cdot,\tilde{\theta}_n)\|^2_{\mathcal{N}_\Phi(\Omega)}\rightarrow \frac{\partial^2}{\partial \theta\partial\theta^\text{T}}\|\epsilon(\cdot,\theta')\|^2_{\mathcal{N}_\Phi(\Omega)},\label{H1}
\end{eqnarray}
because $\tilde{\theta}_n$ tends to $\theta'$. By the definition of $\theta'$ in (\ref{thetaprime}) and the assumption that $\theta'$ is an interior point of $\Theta$, $\frac{\partial}{\partial \theta_i}\|\epsilon(\cdot,\theta')\|^2_{\mathcal{N}_\Phi(\Omega)}=0$. Therefore, we have
\begin{eqnarray}
\frac{\partial}{\partial \theta_i}\|\hat{\epsilon}_n(\cdot,\theta')\|^2_{\mathcal{N}_\Phi(\Omega)}&=& \frac{\partial}{\partial \theta_i}\|\hat{\epsilon}_n(\cdot,\theta')-\epsilon(\cdot,\theta')\|^2_{\mathcal{N}_\Phi(\Omega)}\nonumber\\
&=& 2 \left\langle\frac{\partial (\hat{\epsilon}_n-\epsilon)}{\partial \theta_i}(\cdot,\theta'),\hat{\epsilon}_n(\cdot,\theta')-\epsilon(\cdot,\theta')\right\rangle_{\mathcal{N}_\Phi(\Omega)}\nonumber\\
&\leq& 2 \left\|\frac{\partial (\hat{\epsilon}_n-\epsilon)}{\partial \theta_i}(\cdot,\theta')\right\|_{\mathcal{N}_\Phi(\Omega)} \|\hat{\epsilon}_n(\cdot,\theta')-\epsilon(\cdot,\theta')\|_{\mathcal{N}_\Phi(\Omega)}\nonumber\\
&\leq& 2 C_\Phi^2 h^{2 k}(\mathcal{D}_n) \|v_{\theta'}\|_{L_2(\Omega)}\|D_i v_{\theta'}\|_{L_2(\Omega)}\rightarrow 0,\label{optrate}
\end{eqnarray}
where the first equality follows from (\ref{norm inequality}); the last inequality follows from (\ref{ineq1}) and (\ref{ineq2}); and the limiting relationship follows from (\ref{supv}) and the fact that $h(\mathcal{D}_n)\rightarrow 0$. Then we obtain the desired result by combining (\ref{error}), (\ref{H1}) and (\ref{optrate}).
\hfill $\square$

\subsection{Proof of Theorem \ref{Th L2}}
First we prove the $L_2$-consistency, i.e., the convergence of $\hat{\theta}_{L_2}(\mathcal{D}_n)$ to $\theta^*$. Because $\hat{\theta}_{L_2}(\mathcal{D}_n)$ minimizes $\|\hat{y}^p_n-y^s(\cdot,\theta)\|^2_{L_2(\Omega)}$ and $\theta^*$ is the unique minimizer of $\|\epsilon(\cdot,\theta)\|^2_{L_2(\Omega)}$,
it suffices to prove that $\|\hat{y}^p-y^s(\cdot,\theta)\|^2_{L_2(\Omega)}$ converges to $\|\epsilon(\cdot,\theta)\|^2_{L_2(\Omega)}$ uniformly with respect to $\theta$. 
Note
\begin{eqnarray}
&&\|\hat{y}^p-y^s(\cdot,\theta)\|^2_{L_2(\Omega)}-\|\epsilon(\cdot,\theta)\|^2_{L_2(\Omega)}\nonumber\\
&=& (\|\hat{y}^p-y^s(\cdot,\theta)\|_{L_2(\Omega)}-\|\epsilon(\cdot,\theta)\|_{L_2(\Omega)})\cdot \nonumber\\ &&(\|\hat{y}^p-y^s(\cdot,\theta)\|_{L_2(\Omega)}+\|\epsilon(\cdot,\theta)\|_{L_2(\Omega)})\nonumber\\
&\leq& \|\hat{y}^p-y^s(\cdot,\theta)-\epsilon(\cdot,\theta)\|_{L_2(\Omega)} (\|\hat{y}^p-y^s(\cdot,\theta)\|_{L_2(\Omega)}+\|\epsilon(\cdot,\theta)\|_{L_2(\Omega)})\nonumber\\
&\leq&\|\hat{y}^p-y^p\|_{L_2(\Omega)}(\|\hat{y}^p\|_{L_2(\Omega)}+\|y^s(\cdot,\theta)\|_{L_2(\Omega)}+\|\epsilon(\cdot,\theta)\|_{L_2(\Omega)}). \label{consistency}
\end{eqnarray}
Because $h(\mathcal{D}_n)\rightarrow 0$, it can be seen from (\ref{ordinary}) that $\sup|\hat{y}^p(x)-y^p(x)|\rightarrow 0$. Together with the compactness of $\Omega$, we have $\|\hat{y}^p-y^p\|_{L_2(\Omega)}=o(1)$ and $\|\hat{y}^p\|_{L_2(\Omega)}=O(1)$. Therefore, the uniform convergence is obtained from (\ref{consistency}), and this leads to the $L_2$-consistency of $\hat{\theta}_{L_2}(\mathcal{D}_n)$.

The convergence rate can be derived by following a similar argument as in Theorem \ref{Th rateKO}. Note that for any $f\in L_2(\Omega)$,
\begin{eqnarray}
\|f\|_{L_2}\leq \sqrt{Vol(\Omega)}\sup_{x\in\Omega}|f(x)|, \label{L2volume}
\end{eqnarray}
where $Vol(\Omega)$ is the volume of $\Omega$.
Because $ \|\hat{\epsilon}_n(\cdot,\theta)\|^2_{L_2(\Omega)}$ is minimized at $\hat{\theta}_{L_2}(\mathcal{D}_n)$, $\hat{\theta}_{L_2}(\mathcal{D}_n)$ tends to $\theta^*$ and $\theta^*$ is an interior point, the Taylor expansion gives
\begin{eqnarray}
0&=&\frac{\partial}{\partial \theta}\|\hat{y}^p-y^s(\cdot,\theta^*)\|^2_{L_2(\Omega)}\nonumber\\&&+\bigg(\frac{\partial^2}{\partial \theta\partial \theta^\text{T}}\|\hat{y}^p-y^s(\cdot,\tilde{\theta}_n)\|^2_{L_2(\Omega)}\bigg)(\hat{\theta}_{L_2}(\mathcal{D}_n)-\theta^*),\label{taylorL2}
\end{eqnarray}
where $\tilde{\theta}_n$ is located between $\theta^*$ and $\hat{\theta}_{L_2}(\mathcal{D}_n)$. 
Since $\hat{y}^p$ is independent of $\theta$, it is easy to see that
\begin{eqnarray*}
\frac{\partial^2}{\partial \theta\partial\theta^\text{T}}\|\hat{y}^p-y^s(\cdot,\tilde{\theta}_n)\|^2_{L_2(\Omega)}\rightarrow \frac{\partial^2}{\partial \theta\partial\theta^\text{T}}\|\epsilon(\cdot,\theta^*)\|^2_{L_2(\Omega)},\label{HL2}
\end{eqnarray*}
which is invertible by the assumption.
Therefore the convergence rate is given by
\begin{eqnarray}
&&\frac{\partial}{\partial \theta_i}\|\hat{y}^p-y^s(\cdot,\theta^*)\|^2_{L_2(\Omega)}\nonumber\\
&=&\frac{\partial}{\partial \theta_i}\|\hat{y}^p-y^s(\cdot,\theta^*)\|^2_{L_2(\Omega)}-\frac{\partial}{\partial \theta_i}\|y^p-y^s(\cdot,\theta^*)\|^2_{L_2(\Omega)}\nonumber\\
&=& 2\left\langle y^p-\hat{y}^p,\frac{\partial y^s}{\partial \theta_i}(\cdot,\theta^*)\right\rangle_{L_2(\Omega)}\label{L2norms}\\
&\leq& 2\sqrt{Vol(\Omega)}C_\Phi h^k(\mathcal{D}_n)\|y^p\|_{\mathcal{N}_\Phi(\Omega)}\left\|\frac{\partial y^s}{\partial \theta_i}(\cdot,\theta^*)\right\|_{L_2(\Omega)},\label{hn}
\end{eqnarray}
where the first equality follows from the definition of $\theta^*$ in (\ref{least distance}) and the fact that $\theta^*$ is an interior point; the inequality follows from Schwarz's inequality, (\ref{L2volume}) and (\ref{ordinary}) in Proposition \ref{prop power function}.
Hence (\ref{L2origin}) is obtained by combining (\ref{taylorL2})-(\ref{hn}).

If there exists $v\in L_2(\Omega)$ such that $y^p(x)=\int_\Omega\Phi(x,t)v(t) d t$, (\ref{improved}) in Proposition \ref{prop power function} can be applied to (\ref{L2norms}), which proves (\ref{L2improved}).
\hfill $\square$

\subsection{Proof of Theorem \ref{Th expensive}}
As is in the proof of Theorem \ref{Th L2}, the convergence of $\theta^*(\mathcal{G}_n)$ to $\theta^*$ is a direct consequence of (\ref{ordinary}). As $ \|y^p(\cdot)-\hat{y}^s_n(\cdot,\theta)\|^2_{L_2(\Omega)}$ is minimized at $\theta^*(\mathcal{G}_n)$, the Taylor expansion gives
\begin{eqnarray*}
0&=&\frac{\partial}{\partial \theta}\|y^p(\cdot)-\hat{y}^s_n(\cdot,\theta^*)\|^2_{L_2(\Omega)}\\&&+\bigg(\frac{\partial^2}{\partial \theta\partial \theta^\text{T}}\|y^p(\cdot)-\hat{y}^s_n(\cdot,\tilde{\theta}_n)\|^2_{L_2(\Omega)}\bigg)(\theta^*(\mathcal{G}_n)-\theta^*),
\end{eqnarray*}
where $\tilde{\theta}_n$ is located between $\theta^*$ and $\hat{\theta}_{L_2}(\mathcal{D}_n)$.
It follows from (\ref{L2volume}) and (\ref{ordinary}) in Proposition \ref{prop power function} with $|\alpha|=0,1,2$ that for all $\theta\in\Theta$,
\begin{eqnarray}
\|(y^s-\hat{y}^s_n)(\cdot,\theta)\|_{L_2(\Omega)}&\leq&\sqrt{Vol(\Omega)}C_{\Psi}h^{k'}(\mathcal{G}_n) \|y^s\|_{\mathcal{N}_{\Psi}(\Omega\times\Theta)},\label{ys1}\\
\left\|\frac{\partial (y^s-\hat{y}^s_n)}{\partial\theta_i}(\cdot,\theta)\right\|_{L_2(\Omega)}&\leq&\sqrt{Vol(\Omega)}C_{\Psi}h^{k'-1}(\mathcal{G}_n) \|y^s\|_{\mathcal{N}_{\Psi}(\Omega\times\Theta)},\label{ys2}\\
\left\|\frac{\partial^2 (y^s-\hat{y}^s_n)}{\partial\theta_i\partial\theta_j}(\cdot,\theta)\right\|_{L_2(\Omega)}&\leq& \sqrt{Vol(\Omega)}C_{\Psi}h^{k'-2}(\mathcal{G}_n) \|y^s\|_{\mathcal{N}_{\Psi}(\Omega\times\Theta)},\nonumber
\end{eqnarray}
for $1\leq i,j\leq q$, which implies
\begin{eqnarray*}
\frac{\partial^2}{\partial \theta\partial \theta^\text{T}}\|y^p(\cdot)-\hat{y}^s_n(\cdot,\tilde{\theta}_n)\|^2_{L_2(\Omega)} \rightarrow \frac{\partial^2}{\partial \theta\partial \theta^\text{T}}\|y^p(\cdot)-y^s(\cdot,\theta^*)\|^2_{L_2(\Omega)},
\end{eqnarray*}
which is invertible by the assumption.
Now as in Theorems \ref{Th rateKO} and \ref{Th L2}, we have
\begin{eqnarray}
&&\frac{\partial}{\partial \theta_i}\|y^p(\cdot)-\hat{y}^s_n(\cdot,\theta^*)\|^2_{L_2(\Omega)}\nonumber\\
&=&\frac{\partial}{\partial \theta_i}\|y^p(\cdot)-\hat{y}^s_n(\cdot,\theta^*)\|^2_{L_2(\Omega)}-\frac{\partial}{\partial \theta_i}\|y^p(\cdot)-y^s(\cdot,\theta^*)\|^2_{L_2(\Omega)}\nonumber\\
&=&-2\left\langle y^p(\cdot)-\hat{y}_n^s(\cdot,\theta^*),\frac{\partial \hat{y}_n^s}{\partial \theta_i}(\cdot,\theta^*)\right\rangle_{L_2(\Omega)}\nonumber\\&&+ 2\left\langle y^p(\cdot)-y^s(\cdot,\theta^*),\frac{\partial y^s}{\partial \theta_i}(\cdot,\theta^*)\right\rangle_{L_2(\Omega)}\nonumber\\
&=&2\left\langle y^p(\cdot)-y^s(\cdot,\theta^*),\frac{\partial (y^s-\hat{y}_n^s)}{\partial \theta_i}(\cdot,\theta^*)\right\rangle_{L_2(\Omega)}\label{expensivel2}\\&&+2\left\langle(\hat{y}_n^s-y^s)(\cdot,\theta^*),\frac{\partial \hat{y}_n^s}{\partial\theta_i}(\cdot,\theta^*)\right\rangle_{L_2(\Omega)}.\nonumber
\end{eqnarray}
Because $h(\mathcal{G}_n)\rightarrow 0$, (\ref{ys1}) and (\ref{ys2}) implies that for sufficiently large $n$,
\begin{eqnarray}
\left\|\frac{\partial \hat{y}_n^s}{\partial \theta_i}(\cdot,\theta^*)\right\|_{L_2(\Omega)}\leq 2 \left\|\frac{\partial y^s}{\partial \theta_i}(\cdot,\theta^*)\right\|_{L_2(\Omega)}.\label{bound1}
\end{eqnarray}
By applying Schwarz's inequality to (\ref{expensivel2}) and using the bounds in (\ref{ys1}), (\ref{ys2}), and (\ref{bound1}), we obtain $\frac{\partial}{\partial \theta_i}\|y^p(\cdot)-\hat{y}^s_n(\cdot,\theta^*)\|^2_{L_2(\Omega)}=O(h^{k'-1}(\mathcal{G}_n))$. This leads to the desired result.
\hfill $\square$

\bibliographystyle{siam}
\bibliography{calibration}

\end{document}